\documentclass{!!Arti}
\usepackage[leqno]{amsmath}
\usepackage[psamsfonts]{amssymb}
\usepackage{amsthm}
\usepackage[mathscr,psamsfonts]{eucal}
\usepackage{cmmib57}
\usepackage{!Style}
\usepackage{!Macros}
\usepackage{amscd}
\usepackage{url}
\usepackage[dvips,%
pdftitle={The geometry of Newton's law and rigid systems},%
pdfauthor={M. Modugno, R. Vitolo},%
pdfkeywords={classical mechanics, rigid system, Newton's law,%
  Riemannian geometry}%
]{hyperref}
\newcommand{\cprime}{\/{\mathsurround=0pt$'$}}

\newcommand{\cen}{_{\rm{cen}}}
\newcommand{\dia}{_{\rm{dia}}}
\newcommand{\rel}{_{\rm{rel}}}
\newcommand{\rig}{_{\rm{rig}}}
\newcommand{\Rig}{_{\rm{Rig}}}
\newcommand{\mul}{_{\rm{mul}}}
\newcommand{\anl}{_{\rm{ang}}}
\newcommand{\rot}{_{\rm{rot}}}

\newcommand{\cst}{_{\rm{con}}}
\title{\bf The geometry of Newton's law
\\
and rigid systems}
\author{Marco Modugno$^1$, Raffaele Vitolo$^2$
\myskip
\\
\fz $^1$Department of Applied Mathematics ``G. Sansone"
\\
\fz Via S. Marta 3, 50139 Florence, Italy
\\
\fz {\tt marco.modugno@unifi.it}
\myskip
\\
$^2$\fz Department of Mathematics ``E. De Giorgi"
\\
\fz via per Arnesano, 73100 Lecce, Italy
\\
\fz {\tt raffaele.vitolo@unile.it}}
\date{}
\edition{\fz Preprint: 2005.11.01. - 07.58.}
\begin{document}
\maketitle
\begin{abstract}
We start by formulating geometrically the Newton's law for a
classical free particle in terms of Riemannian geometry, as pattern
for subsequent developments.
In fact, we use this scheme for further generalisation devoted to a
constrained particle, to a discrete system of several free and
constrained particles.

For constrained systems we have intrinsic and extrinsic viewpoints,
with respect to the environmental space.
In the second case, we obtain an explicit formula for the reaction
force via the second fundamental form of the constrained
configuration space.
For multi--particle systems we describe geometrically the splitting
related to the center of mass and relative velocities; in this way we
emphasise the geometric source of classical formulas.

Then, the above scheme is applied in detail to discrete rigid systems.
We start by analysing the geometry of the rigid configuration space.
In this way we recover the classical formula for the velocity of the
rigid system via the parallelisation of Lie groups. Moreover, we
study in detail the splitting of the tangent and cotangent
environmental space into the three components of center of mass, of
relative velocities and of the orthogonal subspace. This splitting
yields the classical components of linear and angular momentum (which
here arise from a purely geometric construction) and, moreover, a
third non standard component.
The third projection yields an explicit formula for the reaction force
in the nodes of the rigid constraint.
\end{abstract}

{\fz {\bf Keywords}:
classical mechanics, rigid system, Newton's law, Riemannian geometry}

{\fz {\bf 2001 MSC}: 70G45, 70B10, 70Exx}

{\fz {\bf Acknowledgements}:
This work has been partially supported by MIUR (Progetto PRIN 2003
``Sistemi integrabili, teorie classiche e quantistiche''), GNFM and
GNSAGA of INdAM, and the Universities of Florence and Lecce.}
\newpage
\tableofcontents
\newpage
\myintro
\label{intro}
The original approach to classical mechanics is based on the Newton's
law.
This is still used and popular mainly in the literature devoted to
applied sciences and engineering, even if it is not very sophisticated
from the mathematical and geometrical viewpoint
(see, for instance, \cite{Gol80,LanLif75,LevAma26,Whi36}).

On the other hand, an approach to mechanics based on modern
differential geometry has been developed and became more and more
popular in the last decades.
This viewpoint is achieved in terms of Riemannian, Lagrangian,
Hamiltonian, symplectic, variational and jet geometry.
A very huge literature exists in this respect (see, for instance,
\cite{AbrMar86,Arn75,Cor91,Cra95,deLRod89,God69,GuiSte84,LibMar87,
Lic62,LitRei97,
MarRat95,MasPag91,MasPag93,ParKim00,Sou69,Tul85,VerFad81}).
These methods have been very successful for
understanding several theoretical aspects and for the solution of
several concrete problems, and have stimulated a large number of
further classical and quantum theories.

\myskip

In this paper, our aims are more specific and foundational.
Namely, we reformulate classical mechanics of a system with a finite
number of particles and rigid systems, in terms of the Newton's law,
in a way which, on one hand, is closer to the classical treatment of
the subjects and, on the other hand, is expressed through the modern
language of differential geometry.

Our approach is addressed both to differential geometers, who could
easily get mechanical concepts written in their language,
and to mathematical physicists, who are interested in a
mathematically rigourous foundation of mechanics.

In fact, several ideas have been achieved independently by
differential geometers and mathematical physicists in different
contexts and with different purposes and languages.
Sometimes, facts which appear in one of the two disciplines as easy
and elementary may correspond to more difficult and
fundamental facts in the other discipline.
We believe that linking those facts provides a new insight on
classical matters and yields new results as well.
For the above reasons, from time to time, we recall some classical
facts of one of the two areas which are possibly not very familiar to
experts of the other area.

Thus, this paper, in spite of the sophisticated mathematical language,
in comparison to the standard literatures of mechanics, analyses
concrete mechanical contents.

On the other hand, this paper provides the classical background for a
covariant approach to the quantisation of a rigid body, which is the
subject of a subsequent paper \cite{ModTejVit05}.

\myskip

The guideline of our approach is the description of mechanics of a
system of $n$ free and constrained particles, including a rigid
system, in terms of the Riemannian formulation of mechanics of one
particle.

We start by recalling the mechanics of one free particle moving in an
affine Euclidean configuration space.  We express the Newton's law in
terms of covariant derivative. In several respects, it is convenient
to introduce forces as forms (instead as vector fields) from the very
beginning.

Then, we can naturally apply this Riemannian approach to the mechanics
of a constrained particle.
We have an intrinsic and an extrinsic viewpoint related to the
embedding of the constrained configuration space into the environmental
space.
In particular, we use the Gauss' Theorem concerning the splitting of
the Riemannian connection in order to get an explicit expression of
the reaction force via the 2nd fundamental form of the constrained
configuration space.

Next, we describe the mechanics of a system of $n$ free particles, as
one free particle moving in a higher dimensional product
configuration space.
For this purpose, it is necessary to introduce a weighted metric
(besides the standard product metric).
Of course, in the case on $n$ free particles, we have the additional
projection on the single particle spaces.
Furthermore, we have the splitting of the configuration space into
the affine component of the center of mass and the vector component
of relative distances.
The 1st splitting can be used to achieve information on the single
particles and is orthogonal with respect to both metrics.
The 2nd splitting has a fundamental role and is orthogonal only
with respect to the weighted metric.
The systematic use of the weighted metric and of the above
orthogonal splitting as a fundamental scheme seems to be original.
In particular, we show that the classical concepts of total
kinetic energy, total kinetic momentum, total force, etc. can be
regarded as a direct consequence of the above geometric scheme.

Then, the formulation of a a constrained system of $n$ particles can
be easily obtained from the above scheme, by repeating the scheme of
one free and constrained particle.

\myskip

Eventually, a particular care is devoted to the analysis of a system
of $n$ particles with a rigid constraint.

First we study the geometry of the rigid configuration space,
distinguishing the non degenerate and degenerate cases.
Then, we formulate the kinematics and mechanics of a rigid system
according to the above scheme.
In particular, we show
(Section \ref{Tangent space of rotational space})
that the classical formula of the velocity of
a rigid system (well--known in mechanics) can be regarded as the
parallelisation of a Lie group (well--known in differential geometry).
This fact yields an interpretation of the inertia tensor as a
representative of the weighted metric induced by the parallelisation.
In this context, we exhibit a new explicit intrinsic expression of the
angular velocity via the inertia tensor
(Corollary \ref{expression of rotational parallelisation}).

By combining the splitting of center of the mass and the splitting of
the constrained configuration space, we obtain a splitting of the
tangent and cotangent environmental configuration spaces into three
components: the component of center of mass, the rotational component
and a further orthogonal component to the configuration space
(Theorem \ref{splitting of the tangent rigid space}
and
Theorem \ref{splitting of the cotangent rigid space}).
This splitting is reflected on all objects of the rigid system
mechanics, providing a clear geometric interpretation of some
classical constructions of mechanics and new results as well.
For instance, the total momentum of forms arises from our
geometric scheme via the projection on the rotational component
(Corollary \ref{total momentum of a multiform}).
Moreover, a special application of the above splitting is the explicit
expression of the reaction force on every node
(Corollary \ref{expression of the rigid reaction}).
This formula seems to be new and possibly useful in engineering
applications.

\myskip

Throughout the paper, we number those formulas which have a key role
and/or emphasise a non standard feature of our approach.

\myskip

We assume all manifolds and maps to be
$\Cin \,.$
If
$\f M$
and
$\f N$
are manifolds, then the sheaf of local smooth maps
$\f M \to \f N$
is denoted by
$\map(\f M, \, \f N) \,.$
\newpage
\section{Preliminaries}
\label{Preliminaries}
\bsm
In this paper we use a few non-standard mathematical constructions.
In order to make the paper self-contained, we start with some
introductory notions.
\esm
\subsection{Scale spaces and units of measurement}
\label{Scale spaces and units of measurement}
\bsm
In order to describe in a rigorous mathematical way the units of
measurements and the coupling scales, we introduce the notion of ``
scale space" \cite{JanModVit05}.
\esm

We define a {\em scale space\/}
$\B U$
as ``positive
$1$-dimensional semi-vector space'' over
$\Rn^+ \,.$
Roughly speaking, this has the same algebraic structure as
$\Rn^+ \,,$
but no distinguished
generator over
$\Rn^+ \,.$
We can naturally define the tensor product between scale spaces
and ordinary vector spaces.
Moreover, we can naturally define the rational powers
$\B U^{p/q}$
of a scale space
$\B U \,.$
Rules analogous to those of real numbers hold for scale spaces;
accordingly, we adopt analogous notation.
In particular, we shall write
$\B U^0 \byd \B U \,,
\;
\B U^{-1} \byd \B U^* \,,
\;
\B U^p \byd \ten^p \B U \,.$

In our theory, these spaces will appear tensorialised with spacetime
tensors. The scale spaces appearing in tensor products are not
effected by differential operators, hence their elements can be
treated as constants.

\myskip

A \emp{coupling scale} is defined to be a scale factor needed for
allowing the equality of two scaled objects and a \emp{unit of
measurements} is defined to be a basis of a scale space.

\myskip

We introduce the scale spaces
$\B T$
of {\em time intervals},
$\B L$
of {\em lengths\/} and
$\B M$
of {\em masses}.

We will consider time units of measurement
$u_0 \in \B T\,,$
or their duals
$u^0 \in \B T^*\,.$
\subsection{Generalised affine spaces}
\label{Generalised affine spaces}
\bsm
Affine spaces are important for classical mechanics because they
offer a geometrical model of the basic configuration spaces.
Moreover, affine spaces constitute the appropriate framework for
elementary differential calculus.  In this paper, we need a more
general definition of the standard notion. Namely, we introduce
generalised affine spaces associated with (possibly non Abelian)
groups. This generalisation is suitable for the description of the
configuration space of rigid systems.
\esm

A \emp{(left) generalised affine space} is defined to be a triple
$(A,\E D A,l) \,,$
where $A$ is a set, $\E DA$ is a group and
$l : \E D A \car A \to A$
is a free and transitive left action.
For the sake of simplicity, we often denote the generalised affine
space
$(A,\E D A,l)$
just by
$A \,.$

For each
$o \in A \,,$
the \emp{left translation}
$l_o : \E D A \to A : g \mto g o$
is invertible.
In fact, for each
$a \in A \,,$
there is a unique
$g \in G \,,$
denoted by
$g \eqv a o^{-1} \,,$
such that
$a = g o \,.$

A \emp{generalised affine map} is defined to be a map
$f : A \to A'$
between generalised affine spaces, such that, for a certain
$o \in A \,,$
we have
$f(a) = \E D f(ao^{-1})f(o) \,,$
for all
$a \in A \,,$
where
$\E D f : \E D A \to \E D A'$
is a group morphism.
We can easily prove that, if such a
$\E D f$
exists, then it is unique and
independent of the choice of
$o \,.$
We say
$\E D f$
to be the \emp{generalised derivative} of
$f \,.$
For example, if
$o \in A \,,$
then the \emp{left translation}
$l_o : \E D A \to A$
is a generalised affine map and its derivative is just the identity.

Of course, if
$A$
is a generalised affine space associated with the additive group of a
vector space
$V \,,$
then
$A$
turns out to be an affine space according to the standard definition
and the notions of generalised affine map and generalised derivative
reduce to the standard ones.

\myskip

Now, let us consider a generalised affine space
$A$
associated with a Lie group
$G \,.$
Then, there is a unique smooth structure of
$A \,,$
such that the left translation
$l : G \car A \to A$
be smooth.

Let
$\F g \byd T_eG$
be the Lie algebra of the group
$G \,,$
where
$e \in G$
is the unit element.

We recall the following well--known result (see, for instance,
\cite{War71}).

\bLm\label{parallelisation of a Lie group}
The Lie group
$G$
is parallelisable through a natural isomorphism
$TG \seq G \car \F g \,.$
\eLm

\bpf
Let us consider the map
$Tl : TG \car TG \to TG$
and the trivial vector subbundles
$T_eG \sub TG \,,$
over
$\{e\} \sub G \,,$
and
$G \sub TG \,,$
over
$G \,.$

Then, the restriction of
$Tl$
to the vector subbundle
$T_eG \car G$
yields, the linear fibred isomorphism
$Tl|_{T_eG \car G} : T_eG \car G \to TG$
over
$G \,.$

Thus, for each
$g \in G \,,$
we have the linear isomorphism
$Tl_g|_{T_eG} : T_eG \to T_gG \,.$\QED
\epf

We can easily generalise the above result to the affine space
$A \,.$

\bLm\label{parallelisation of an affine space}
The affine space
$A$
is parallelisable through a natural isomorphism
$TA \seq A \car \F g \,.$
\eLm

\bpf
Let us consider the map
$Tl : TG \car TA \to TA$
and the trivial vector subbundles
$T_eG \sub TG \,,$
over
$\{e\} \sub G \,,$
and
$A \sub TA \,,$
over
$A \,.$

Then, the restriction of
$Tl$
to the vector subbundle
$T_eG \car A$
yields, the linear fibred isomorphism
$Tl|_{T_eG \car A} : T_eG \car A \to TA$
over
$A \,.$

Thus, for each
$a \in A \,,$
we have the linear isomorphism
$Tl_a|_{T_eG} : T_eG \to T_aA \,.$\QED
\epf

\bLm\label{connection induced by a parallelisation}
Let
$\f M$
be a differential manifold equipped with a parallelisation
$T\f M \seq \f M \car \f F \,,$
which induces a projection
$TT\f M \to T\f M \,.$
Then, we obtain the linear connection, given for each sections
$X, Y : \f M \to T\f M \,,$
by means of the composition
\bcd
\f M @>{\nab_Y X}>> T\f M
\\
@V{Y}VV
@AAA
\\
T\f M @>{TX}>> TT\f M \,.\END
\ecd
\eLm

Thus, the parallelisations of
$G$
and
$A$
induce linear connections of
$G$
and
$A \,.$
\newpage
\section{Mechanics of one particle}
\label{Mechanics of one particle}
\bsm
First, we review the one free and constrained particle mechanics as an
introduction to our formalism and a pattern for next generalisations.
\esm
\subsection{Free particle}
\label{Free particle}
\bsm
We start with a free particle moving in a 3-dimensional Euclidean
affine space
$\f P \,.$
On the other hand, in many respects, the dimension 3 and the affine
structure have no essential role; in fact, we essentially exploit the
underlying weaker structure of Riemannian manifold of
$\f P \,.$
\esm
\subsubsection{Configuration space}
\label{Free particle: Configuration space}
We define the \emp{time} to be a 1-dimensional affine space
$\f T$
associated with the vector space
$\baB T \byd \B T \ten \Rn \,.$
We shall always refer to an affine chart $(x^0)$ induced by an origin
$t_0\in\f T$
and a time unit of measurement
$u^0 \in \B T \,.$

We define the \emp{pattern configuration space} to be a
3-dimensional affine space
$\f P$
associated with an oriented vector space
$\f S \,.$
We shall refer
to a (local) chart
$(x^i)$
on
$\f P \,.$
Latin indices
$i \,, j \,, h \,, k$
will run from
$1$ to $3 \,.$

We shall also be involved with the tangent space
$T\f P =\f P \car \f S$
and the cotangent space
$T^*\f P = \f P \car \f S^* \,.$
We shall refer to the local charts
$(x^i,\dot x^i)$
of
$T\f P$
and
$(x^i,\dot x_i)$
of
$T^*\f P$
and to the corresponding local bases of vector fields
$\der_i$
and forms
$d^i \,.$
We also denote by
$(x^i, \dt x^i, \pr x^i, \ddt x^i)$
the induced chart of
$TT\f P \,,$
with
$(\der_i, \dt\der_i)$
and
$(d^i,\dot d^i)$
the corresponding bases of vector fields and 1-forms; we have the
chart
$(x^i,\dot x^i,\ddot x^i)$
of
$VT\f P \,.$

The parallelisation of
$\f P$
induced by the affine structure yields a flat linear connection
$\nab$
(see Lemma \ref{connection induced by a parallelisation}.)

We equip
$\f S$
with a scaled Euclidean metric
$g \in \B L^2 \ten (\f S^* \ten \f S^*) \,,$
called \emp{pattern metric}, which can be regarded as a scaled
Riemannian metric of
$\f P$
\[
g : \f P \to
\B L^2 \ten (\f T^*\f P \ten \f T^*\f P) \,.
\]

We denote by
$\ba g$
the corresponding contravariant metric.
We have the coordinate expressions
$g = g_{ij} \, d^i\ten d^j$
and
$\ba g = g^{ij} \, \der_i \ten \der_j \,,$
with
$g_{ij} \in \map(\f P, \, \B L^2 \ten \Rn)$
and
$g^{ij} \in \map(\f P, \, \B L^{-2} \ten \Rn) \,.$
The associated \emp{flat isomorphism} and its inverse,
the \emp{sharp isomorphism}, are denoted by
$g\Fla :T\f P \to \B L^2 \ten T^*\f P$
and
$g\Sha : T^*\f P \to \B L^{-2} \ten T\f P \,.$
The metric
$g$
and an orientation  of
$\f S$
yield the scaled volume form
$\eta \in \B L^3 \ten \Lam^3\f S^*$
and its inverse
$\ba\eta \in \B L^{*3} \ten \Lam^3\f S \,.$

\myskip

The Riemannian connection associated with
$g$
coincides with
$\nab \,.$
We denote the vertical projection associated with
$\nab$
by
$\nu : TT\f P \to T\f P$
and the the Christoffel symbols by
$\Gam^i_{hk} \,.$
\subsubsection{Kinematics}
\label{Free particle: Kinematics}
We define the \emp{phase space} as the 1st jet space of maps
$\f T \to \f P$
\[
J_1\f P \byd \f T \car (\B T^{-1} \ten T\f P)
= \f T \car \f P \car
(\B T^{-1} \ten \f S)\,.
\]

The induced chart of
$J_1\f P$
is
$(x^0, \, x^i, \, x^i_0) \,.$

\myskip

A \emp{motion} is defined to be a map
$s : \f T \to \f P \,.$

The \emp{1st differential}, the \emp{2nd differential}, the
\emp{velocity} and the \emp{acceleration} of a motion
$s$
are defined to be, respectively, the maps
\bgt
ds :
\f T \to \B T^{-1} \ten T\f P \,,
\qquad
d^2s :
\f T \to \B T^{-1} \ten T(\B T^{-1} \ten T\f P) \,,
\qquad
j_1s :
\f T \to J_1\f P \,,
\\
\nab ds :
\f T \to \B T^{-2} \ten T\f P \,.
\end{gather*}

By definition, we have
$j_1s(t) = (t, ds(t))$
and
$\nab ds = \nu \com d^2s \,.$

Moreover, by taking into account the splittings
\bal
\B T^{-1} \ten T\f P
&\seq
\f P \car (\B T^{-1} \ten \f S) \,,
\\
\B T^{-1} \ten T(\B T^{-1} \ten T\f P)
&\seq
\big(\f P \car (\B T^{-1} \ten \f S)\big) \car
\big((\B T^{-1} \ten \f S) \car (\B T^{-2} \ten \f S)\big) \,,
\end{align*}
we can write
\[ds = (s, Ds) \,,
\qquad
d^2s = (s, Ds, Ds, D^2s) \,,
\qquad
\nab ds = (s, D^2s) \,,
\]
where
$Ds : \f T \to \B T^{-1}\ten \f S$
is the standard derivative of
$s \,.$

We have the coordinate expressions
\bal
(x^i, \dot x^i_0) \com ds
&=
(s^i, \, \der_0 s^i) \,,
\\
(x^i, \dot x^i_0, \pr x^i_0, \ddt x^i_{00}) \com d^2s
&=
(s^i, \, \der_0 s^i, \, \der_0 s^i, \, \der_0^2 s^i) \,,
\\
(x^0, x^i, x^i_0) \com j_1s
&=
(x^0, \, s^i, \, \der_0 s^i) \,,
\\
(x^i, \ddt x^i_{00}) \circ \nab ds
&=
(s^i, \,
\der_0^2 s^i +
(\Gam^i_{hk} \com s) \, \der_0 s^h \, \der_0 s^k) \,.
\end{align*}

\myskip

With reference to a mass
$m \in \B M \,,$
we define the \emp{kinetic energy}
and the
\emp{kinetic momentum}, respectively, to be the maps
\bat{2}
\C K &:
\B T^{-1} \ten T\f P \to (\B T^{-2} \ten \B L^2 \ten \B M) \ten
\Rn
&&:
v \mto \tfr12 m \, g(v,v)
\\
\C P \byd D\C K
&:
\B T^{-1} \ten T\f P \to (\B T^{-1} \ten \B L^2 \ten \B M)
\ten T^*\f P
&&:
v \mto m \, g\Fla(v) \,.
\end{alignat*}

We have the coordinate expressions
$\C K = \tfr12 m \, g_{ij} \,\dt x^i_0 \, \dt x^j_0$
and
$\C P = m \, g_{ij} \, \dt x^i_0 \, d^j \,.$
\subsubsection{Dynamics}
\label{Free particle: Dynamics}
In our context, the force acting on a particle is given a priori on
the the phase space. Moreover, it is convenient to introduce the
force as a co--vector.

Thus, a \emp{force} is defined to be a map
\[
F :
J_1\f P \to (\B T^{-2} \ten \B L^2 \ten \B M) \ten T^*\f P \,.
\]

The force
$F$
is said to be \emp{conservative} if it factorises through
$J_1\f P \to \f P$
and can be derived from a
\emp{potential}
$\C U : \f P \to \B T^{-2} \ten \B L^2 \ten \B M \ten \Rn$
by the equality
$F = d\C U \,.$
If the force is conservative, then we define the associated
\emp{Lagrangian} to be the map
$\C L \byd \C K + \C U \,.$

We say that a motion $s$ fulfills the \emp{Newton's law of
motion} if
\bEq
m \, g\Fla \, (\nab ds) = F  \com j_1s \,.
\eEq

It is remarkable that we can link the formulation of dynamics in terms
of the connection
$\nab$
with the Lagrangian approach, directly without any reference to
variational or Lagrangian calculus.
In fact, the following \emp{Lagrange's formula} holds
\[
m \, g\Fla \ (\nab ds) =
\big(
D(\dot\der_i \C K \com ds) - \der_i\C K \com ds
\big) \, (d^i \com s) \,.
\]
By the way, the above formula provides quickly the Christoffel's
symbols of
$\nab \,.$

Hence, the coordinate expression of the Newton's law is
\[
m \, g_{ij} \, \big(
\der_0^2 s^j +
(\Gam^j_{hk} \com s) \, \der_0 s^h \, \der_0 s^k\big)
\eqv
D(\dot\der_i \C K \com ds) - \der_i\C K \com ds =
F_i \com j_1s \,.
\]
In the particular case when the force is conservative, the Newton's
law of motion is expressed by the \emp{Lagrange equations}
\[
D(\dt\der_i \C L \com ds) - \der_i \C L \com ds = 0\,.
\]
\subsection{Constrained particle}
\label{Constrained particle}
\bsm
We assume an embedded submanifold of the pattern Eu\-clidean affine
space as configuration space of a constrained particle.

The mechanics of a constrained particle has two features: an
\emp{intrinsic} and an \emp{extrinsic} one.
According to the intrinsic viewpoint, the particle behaves as a
`free' particle moving in an $l$-dimensional Riemannian manifold;
hence, according to the intrinsic viewpoint, we can repeat the scheme
of the previous section.  On the other hand, the environment space
adds an exterior geometric structure: the 2nd fundamental form, which
measures the deviation of the submanifold from being an affine
subspace of the environmental space.
Then, according to the extrinsic viewpoint, we interpret the reaction
force in terms of the 2nd fundamental form of the constrained
space.
\esm
\subsubsection{Configuration space}
\label{Constrained particle: Configuration space}
We define the \emp{configuration space} for a constrained
particle to be an embedded submanifold of dimension $1\leq l\leq 3$
\[
i\cst : \f P\cst\hto \f P\,.
\]

Thus, by definition of embedded submanifold, for each
$p\in \f P\cst \,,$
there exists a chart
$(x^i)$
of
$\f P$
in a neighbourhood of
$p \,,$
such that
$\f P\cst$
is locally characterised by the constraint
$\{x^{l+1} = 0, \dots, x^3 = 0\} \,.$
Then,
$(y^1, \dots, y^l) \byd (x^1|_{\f P\cst}, \dots, x^l|_{\f P\cst})$
turns out to be a local chart of
$\f P\cst \,.$
The functions
$y^1, \dots, y^l$
are said to be local \emp{Lagrangian coordinates} and the functions
$x^{l+1}, \dots, x^3$
to be local \emp{constraints}.
 From now on, we shall refer to such adapted charts.

For practical reasons, we shall adopt the following convention:

- indices $i,j,h,k$ will run from $1$ to $3 \,;$

- indices $a, b, c, d$ will run from $1$ to $l \,;$

- indices $r, s, t$ will run from $l$ to $3 \,.$

We have
$T\f P|_{\f P\cst} = \f P\cst \car \f S$
and
$T^*\f P|_{\f P\cst} = \f P\cst \car \f S^* \,.$

We have the natural dual linear injective and projective maps
\[
Ti\cst : T\f P\cst \hto T\f P|_{\f P\cst} \sub T\f P
\ssep{and}
\pi \byd T^*i\cst : T^*\f P|_{\f P\cst} \to T^*\f P\cst \,,
\]
with coordinate expressions
\[
Ti\cst (\sum_{1 \leq a \leq l} X^a \, \der_a) =
\sum_{1 \leq a \leq l} X^a \, \der_a
\ssep{and}
\pi(\sum_{1 \leq i \leq n} \ome_i \, d^i) =
\sum_{1 \leq a \leq l} \ome_a \, d^a \,.
\]

The complementary linear projective and injective maps can be
obtained by means of the metric.

We consider the orthogonal subspaces
\bat{2}
T\Per \f P\cst
&\byd
\{X \in T\f P |_{\f P\cst} \mid
g(X, \, T\f P\cst) = 0\}
&&\sub
T\f P|_{\f P\cst} \,,
\\
T\per \f P\cst
&\byd
\{\alp \in T^*\f P |_{\f P\cst} \mid
\alp(T\f P\cst) = 0\}
&&\sub
T^*\f P|_{\f P\cst} \,.
\end{alignat*}

The vector fields
$\der_a$
are tangent to
$\f P\cst \,,$
while the vector fields
$\der_r$
are transversal.
If
$\der_r \in T\Per \f P\cst \,,$
then the adapted chart
$(x^i)$
is said to be \emp{orthogonal} to the submanifold.

The subspace
$T\per\f P\cst$
consists of the forms of the type
$\ome = \sum_{l+1 \leq r \leq n} \ome_r \, d^r \,,$
i.e. of forms whose ``tangent" components vanish.

The restriction
\[
g\cst \byd i\cst^*g :
\f P\cst \to \B L^2 \ten (T^*\f P\cst \ten T^*\f P\cst)
\]
of the pattern metric
$g$
to
$\f P\cst$
is a scaled Riemannian metric, which will be called the
\emp{intrinsic metric}.
Its coordinate expression is
$g\cst = \sum_{1 \leq a,b \leq l} (g\cst)_{ab}\,d^a \ten d^b \,,$
where we have set
$(g\cst)_{ab} \byd g_{ab}|_{\f P\cst} \,.$
The contravariant form of
$g\cst$ will be denoted by
$\ba g\cst \,.$
We stress that, in general, the
$l \car l$
``tangent" submatrix of
$(g^{ij})$
is different from the inverse of the matrix
$(g_{ab}) \,;$
they are equal if and only if the adapted chart
$(x^i)$
is orthogonal.

The intrinsic metric
$g\cst$
yields the Riemannian connection
$\nab\cst \,.$

With reference to a mass
$m \in \B M \,,$
we define the \emp{intrinsic kinetic energy} and the
\emp{intrinsic kinetic momentum}
\bat{2}
\C K\cst
&\byd
i^* \C K :
\B T^{-1} \ten T\f P\cst \to
(\B T^{-1} \ten \B L^2 \ten \B M) \ten \Rn
&&:
v \mto \tfr12 m \, g\cst(v,v) \,,
\\
\C P\cst
&\byd
i^* \C P :
\B T^{-1} \ten T\f P\cst \to
(\B T^{-1} \ten \B L^2 \ten \B M) \ten T^*\f P
&&:
v \mto m \, g\Fla\cst (v) \,,
\end{alignat*}
with coordinate expressions
$\C K\cst = \tfr12 m \, g\cst\,_{ab} \, \dt y^a \, \dt y^b$
and
$\C P\cst = m \, g\cst\,_{ab} \, \dt y^a \, d^b \,.$

\myskip

Let us analyse the orthogonal splittings of the tangent and cotangent
spaces induced by the metric.

The metric yields the injective map
\[
\jmath : T^*\f P\cst \hto T^*\f P|_{\f P\cst}
\]
through the commutative diagram
\bcd
T^*\f P\cst
@>{\jmath} >> T^* \f P|_{\f P\cst}
\\
@V{g\Sha\cst}VV  @AA{\pi}A
\\
\B L^{-2} \ten T\f P\cst
@>{g\Fla} >>  \B L^{-2} \ten T^*\f P|_{\f P\cst}
\ecd

We have the mutually dual orthogonal splittings
\bal
T\f P|_{\f P\cst}
&=
T\f P\cst \udrs{\f P\cst} T\Per \f P\cst \,,
\\
T^*\f P|_{\f P\cst}
&=
T^*\f P\cst \udrs{\f P\cst} T\per \f P\cst \,,
\end{align*}
with projections
\bat{4}
\pi\Prl
&: T\f P|_{\f P\cst}
&&\to
T\f P\cst \,,
&&\qquad
\pi\Per : T\f P|_{\f P\cst}
&&\to
T\Per \f P\cst
\\
\pi
&: T^*\f P|_{\f P\cst}
&&\to
T^*\f P\cst \,,
&&\qquad
\pi\per:T^*\f P|_{\f P\cst}
&&\to
T\per \f P\cst \,.
\end{alignat*}

As the projection
$\pi$
has a very simple expression, it is convenient to compute the other
projections
$\pi\Prl, \, \pi\Per, \, \pi\per$
via the following commutative diagrams
\bcd
T\f P|_{\f P\cst}
@>{\pi^{\|}} >> T \f P\cst
\\
@V{g\Fla}VV
@AA{g\Sha\cst}A
\\
\B L^2 \ten T^*\f P|_{\f P\cst}
@>{\pi} >> \B L^2 \ten T^*\f P\cst
\ecd
\bcd
T^*\f P|_{\f P\cst}
@>{\pi\per} >> T\per \f P\cst
\\
@V{g\Sha}VV  @AA{g\Per{\Fla}}A
\\
\B L^{-2} \ten T\f P|_{\f P\cst}
@>{\pi\Per} >>  \B L^{-2} \ten T\Per \f P\cst
\ecd

Then, for each
$X\in T\f P|_{\f P\cst}$
and
$\ome \in T^*\f P|_{\f P\cst} \,,$
we obtain the equalities
\bal
\jmath(\ome)
&=
\sum_{1 \leq a,b,c \leq l} g_{cb} \, g^{ab} \, \ome_a \, d^c
\,,
\\
\pi^{\|}(X)
&=
\sum_{1 \leq i \leq n} \, \, \sum_{1 \leq a,b \leq l}
X^i \, g_{ib} \, g\cst^{ba} \, \der_a
=
\sum_{1 \leq a,b \leq l} \, \, \sum_{l+1 \leq r \leq n}
(X^a + X^r \, g_{rb} \, g\cst^{ba}) \, \der_a \,,
\\
\pi\Per(X)
&=
\sum_{1 \leq a,b \leq l} \, \, \sum_{l+11 \leq r \leq l}
X^r (\der_r - g_{rb} \, g\cst^{ba} \, \der_a) \,,
\\
\pi\per (\ome)
&=
\sum_{1 \leq a,b \leq l} \, \, \sum_{l+1 \leq r \leq n}
(\ome_r - \ome_a \, g\cst^{ab} \, g_{br}) \, d^r \,.
\end{align*}

Of course, the above formulas simplify considerably if
the adapted chart is orthogonal, i.e. if
$g_{rb}|_{\f P\cst} = 0 \,.$
\subsubsection{Kinematics}
\label{Constrained particle: Kinematics}
We define the \emp{intrinsic phase space} as the 1st jet space of
maps
$\f T \to \f P\cst$
\[
J_1\f P\cst \byd \f T \car (\B T^{-1} \ten T\f P\cst)\,.
\]

The induced chart of
$J_1\f P\cst$
is
$(x^0, \, y^a, \, y^a_0) \,.$

\myskip

A \emp{constrained motion} is defined to be a map
$s\cst : \f T \to \f P\cst \sub \f P \,.$
Clearly, a constrained motion can be naturally regarded as a motion of
the pattern space, via the inclusion
$i\cst \,.$
Indeed, a motion
$s : \f T \to \f P$
is constrained if and only if
$s^r = 0 \,.$

The \emp{1st differential}, the \emp{2nd differential} and the
\emp{velocity} of a constrained motion
$s\cst \,,$
computed in the environment space, turn out to be valued in the
corresponding constrained subspaces
\bgt
ds\cst :
\f T \to \B T^{-1} \ten T\f P\cst \,,
\qquad
d^2s\cst :
\f T \to \B T^{-1} \ten T(\B T^{-1} \ten T\f P\cst) \,,
\\
j_1s\cst :
\f T \to J_1\f P\cst \,.
\end{gather*}

We have the coordinate expressions
\bal
(y^a, \dt y^a_0) \com ds\cst
&=
(s^a, \, \der_0 s^a) \,,
\\
(y^a, \dt y^a_0, \pr y^a_0, \ddt y^a_{00}) \com d^2s\cst
&=
(s^a, \, \der_0 s^a, \, \der_0 s^a, \, \der_0^2 s^a)
\\
(x^0, y^a, y^a_0) \com j_1s\cst
&=
(x^0, \, s^a, \, \der_0 s^a) \,.
\end{align*}

Hence, as far as the above objects are considered, the intrinsic and
the extrinsic approaches coincide, up to the natural inclusion of the
constrained spaces into the corresponding environmental spaces.

Conversely, the intrinsic and extrinsic approaches of the acceleration
of the constrained motion
$s\cst$
do not coincide.
The intrinsic viewpoint is suitable for the intrinsic expression
of the law of motion and the extrinsic viewpoint provides the
constraint reaction force.

We define the \emp{intrinsic acceleration} of a constrained motion
$s\cst$
as the map
\[
\nab\cst ds\cst : \f T \to \B T^{-2} \ten T\f P\cst \,,
\]
with coordinate expression
\[
\nab\cst ds\cst = \sum_{1 \leq a,b,c \leq l}
\big(\der_0^2s^a + (\Gam\cst\,^a_{bc})
\com s\cst \, \der_0s^b \, \der_0s^c \big) (\der_a \com s\cst) \,.
\]

Analogously to the free case, the intrinsic co-acceleration is given
by the Lagrange's formula
\[
m \, g\cst{\Fla} (\nab\cst ds) =
\big(D (\dt\der_a \C K\cst \com ds\cst) -
\dt\der_a \C K\cst \com ds\cst\big) (d^a \com s\cst) \,.
\]

On the other hand, by regarding the constrained motion
$s\cst$
as a motion of the environmental space, we define the \emp{extrinsic
acceleration} as the map
\[
\nab ds\cst : \f T \to \B T^{-2} \ten T\f P \,,
\]
with coordinate expression
\[
\nab ds\cst = \sum^{1 \leq i \leq 3}_{1 \leq a,b,c \leq l}
\big(\der_0^2s^i + (\Gam^i_{bc})
\com s\cst \, \der_0s^b \, \der_0s^c \big) (\der_i\com s\cst) \,.
\]

Then, according to the Gauss' Theorem
\cite{GalHulLaf90},
we have the splitting
\[
\nab ds\cst = \pi\Prl(\nab ds\cst) + \pi\Per(\nab ds\cst) \,,
\]
with
\[
\pi\Prl(\nab ds\cst) = \nab\cst ds\cst
\ssep{and}
\pi\Per \com \nab ds\cst = N \com ds\cst \,,
\]
where
\[
N : T\f P|_{\f P\cst} \to T\Per \f P\cst
\]
is a quadratic map, called \emp{2nd fundamental form}, whose
coordinate expression is
\[
N = \sum_{1 \leq a,b,c \leq l} \, \, \sum_{l + 1 \leq r \leq 3}
\Gam_{bc}^r \, \dt y^b \, \dt y^c \,
(\der_r - g_{rb} \, g\cst^{ba} \der_a) \,.
\]

Indeed, the map $N$ measures how the submanifold
$\f P\cst$
deviates, at 1st order, from being an affine subspace of
$\f P \,.$
The quickest way to compute the 2nd fundamental form is the following:
compute the covariant expressions of the extrinsic and intrinsic
acce\-lerations via the Lagrange's formulas; then pass to the
contravariant expressions and take the difference.
\subsubsection{Dynamics}
\label{Constrained particle: Dynamics}
Let us consider a force
$\ti F : J_1\f P \to
(\B T^{-2} \ten \B L^2 \ten \B M) \ten T^*\f P$
in the environment space
$\f P \,.$
As we are dealing with constrained mechanics, we are involved only
with its restriction
\[
F \byd \ti F|_{J_1\f P\cst} : J_1\f P\cst \to
(\B T^{-2} \ten \B L^2 \ten \B M) \ten T^*\f P|_{\f P\cst} \,.
\]

According to the splitting of
$T^*\f P|_{\f P\cst} \,,$
we can write
\[
F = F\cst + F\cst\,\per \,,
\]
where
\[
F\cst = i^* F :
J_1\f P\cst \to (\B T^{-2} \ten \B L^2 \ten \B M) \ten T^*\f P\cst \,.
\]
We call
$F\cst$
the \emp{intrinsic force}.

\myskip

Let us assume that constraint confines the motion on the configuration
space
$\f P\cst \,,$
via Newton's law of motion, by means of a suitable additional
`reaction force' defined on the constrained space
\[
R : J_1\f P\cst \to
(\B T^{-2} \ten \B L^2 \ten \B M) \ten T^*\f P|_{\f P\cst} \,.
\]

According to the splitting of
$T^*\f P|_{\f P\cst} \,,$
we can write
\[
R = R\cst + R\cst\,\per \,,
\]
where
\[
R\cst = i^* R :
J_1\f P\cst \to (\B T^{-2} \ten \B L^2 \ten \B M) \ten T^*\f P\cst \,.
\]
We call
$R\cst$
the \emp{intrinsic reaction force}.

A constrained motion
$s\cst : \f T\to \f P\cst$
is said to fulfill the \emp{constrained Newton's law of motion} if the
following equation holds
\[
m \, g\Fla \com (\nab ds\cst) = (F + R) \com j_1s\cst \,.
\]

\bTh\label{constrained particle: splitting of the Newton's law}
A constrained motion
$s : \f T\to \f P\cst$
fulfills the \emp{constrained Newton's law of motion} if and only if
\bal
m \, g\cst\Fla \com (\nab\cst ds\cst)
&=
(F\cst + R\cst) \com j_1s\cst
\\
m \, g\Fla \com (N\com ds\cst)
&=
(F\cst\,\per + R\cst\,\per) \com j_1s\cst \,.
\end{align*}

Actually, for each choice of initial data in
$J_1\f P\cst \,,$
the 1st equation has (locally) a unique solution and the 2nd equation
is fulfilled if and only if
\bEq
R\cst\,\per = m \, g\Fla \com N - F\cst\,\per \,.\END
\eEq
\eTh

According to the above result, we make the ``minimal assumption"
(\emp{virtual works principle}, i.e., \emp{smooth constraint})
\[
R\cst = 0 \,.
\]

Then, the explicit coordinate expression of
$R$
is
\bEq
R = \sum^{l+1 \leq r \leq 3}_{1 \leq a,b,c \leq l}
\left(m \, (\Gam_{arb} -
g_{rc} \, g\cst^{cd} \, \Gam\cst\,_{adb}) \,
\dt y^a \, \dt y^b
- F_r + g\cst^{ba} \, F_a \, g_{br} \right) \com i\cst \,
(d^r \com i\cst) \,.
\eEq

In classical literature (see, for instance, \cite{LevAma26}) the
computation of reaction force is presented implicitly as the solution
of a linear system associated with Lagrange multipliers. Instead, the
above formula (which involves an adapted chart) provides an explicit
expression of the reaction force in terms of the Christoffel symbols,
or of the metric.
In the particular case when the adapted chart is orthogonal, this
formula becomes very easy.

\myskip

Thus, the dynamics of a constrained particle can be interpreted as the
dynamics of a `free' particle moving on the Riemannian configuration
space
$\f P\cst \,.$
Then, all main notions and results holding in the free case by using
the Riemannian structure of
$\f P$
can be easily rephrased in the constrained case.
This is a remarkable conceptual and practical advantage of the present
approach.
\vfill
\newpage
\section{Mechanics of a system of $n$ particles}
\label{Mechanics of a system of $n$ particles}
\bsm
In this section we generalise the previous concepts and results to
systems of many particles. Our guideline will be the
interpretation of the multi-particle system as a one-particle
moving in a higher dimensional space.
In this way, all we have learned for one-particle can be applied
directly to multi-particle systems.
On the other hand, we have additional concepts, e.g. the center of
mass splitting, which follow from the projections on the factor spaces
of the different particles.
\esm
\subsection{Free particles}
\label{Free particles}
\bsm
We start with a finite system of free particles.

The key feature of our analysis is the geometry of the configuration
space, which is naturally equipped with two metrics and two natural
splittings.
\esm

We shall use systematically the prefix ``multi" to indicate objects
of the $n$--system analogous to objects of one-particle.

We assume
$n\ge 1 \,,$
and consider $n$ masses
$m_1 \,, \dots,\, m_n \in \B M \,.$

For each
$i = 1, \dots, n \,,$
with reference to the $i$--th particle, it is convenient to consider
a copy of the following pattern objects:
$\f P_i\eqv \f P$ and
$\f S_i\eqv \f S \,.$
\subsubsection{Geometry of the multi--configuration space}
\label{Geometry of the multi--configuration space}
The \emp{multi-configuration space} is defined to be the product
space
\[
\f P\mul \byd \f P_1 \car \cdots \car \f P_n \,.
\]

Clearly,
$\f P\mul$
is an affine space associated with the vector space
$\f S\mul \byd \f S_1\car \cdots \car \f S_n \,.$

A product chart
$(x^i\mul)$
of
$\f P\mul$
induced by charts
$(x^i_1) \,, \dots \,, (x^i_n)$
of the single components is said to be \emp{without interference of
the particles}.
Conversely, a chart
$(x^i\mul)$
of
$\f P\mul$
which cannot be written as a product as above is said to be
\emp{with interference of the particles}.

\myskip

We define the \emp{total mass} as
$m_0 \byd \sum_i m_i \in \B M$
and the $i$-th
\emp{weight} as
$\mu_i = m_i/m_0 \,.$
Clearly, we have
$\sum_i \mu_i = 1 \,.$

A typical notation for the elements of
$\f P\mul \,,$
$\f S\mul$
and
$\f S\mul^*$
will be
\bat{3}
p\mul
&=
(p_1, \dots, p_n)
&&\in
\f P\mul
&&=
\f P_1 \car \dots \car \f P_n \,,
\\
v\mul
&=
(v_1, \dots, v_n)
&&\in
\f S\mul
&&=
\f S_1 \car \dots \car \f S_n \,,
\\
\alp\mul
&=
(\alp_1, \dots, \alp_n)
&&\in
\f S\mul^*
&&=
\f S^*_1 \car \dots \car \f S^*_n \,.
\end{alignat*}

We define the \emp{multi-geometrical metric} and the
\emp{multi-weighted metric} as
\bAt{2}
g\mul
&:
\f P\mul \to \B L^2 \ten (T^*\f P\mul \ten T^*\f P\mul)
&&:
(u\mul,v\mul) \mto \sum_i g(u_i,v_i) \,;
\\
G\mul
&:
\f P\mul \to \B L^2 \ten (T^*\f P\mul \ten T^*\f P\mul)
&&:
(u\mul,v\mul) \mto \sum_i \mu_i \, g(u_i,v_i) \,.
\end{alignat}

The contravariant tensors of
$g\mul$
and
$G\mul$
are denoted, respectively, by
$\ba g\mul$
and
$\ba G\mul \,.$

If
$n=1 \,,$
then
$G\mul = g\mul = g$
and
$m = m_0 \,.$

If
$n\ge 2 \,,$
then the two metrics
$G\mul$
and
$g\mul$
are distinct.
Moreover, if
$m_1 = \dots = m_n \,,$
then
$\mu_1 = \dots = \mu_n = \tfr1n$
and
$G\mul = \tfr1n g\mul \,.$

\myskip

For a system of $n$ particles, we will rephrase the dynamics of a
system of one particle, by replacing the pattern metric
$g$
with the weighted metric
$G\mul$
and the mass
$m$
with the total mass
$m_0 \,.$
This procedure yields the correct Newton's law of motion, in full
analogy with the one particle case.

According to this scheme, we define the \emp{multi-kinetic energy} and
the \emp{multi-kinetic momentum} by
\bat{2}
\C K\mul
&:
\B T^{-1} \ten T\f P\mul \to
(\B T^{-1} \ten \B L^2 \ten \B M) \ten \Rn
&&: v \mto \tfr12 m_0 \, G\mul(v,v) \,,
\\
\C P\mul
&:
\B T^{-1} \ten T\f P\mul \to
(\B T^{-1} \ten \B L^2 \ten \B M) \ten T^*\f P\mul
&&: v \mto m_0 \, G\Fla\mul (v) \,,
\end{alignat*}
and recover the standard formulas
$\C K\mul(v) =  \sum_i \tfr12 m_i \, g(v_i, v_i)$
and
$\C P\mul (v) = \big(m_i \, g(v_i)\big) \,.$

The linear connection
$\nab\mul$
induced on the multi-configuration space
$\f P\mul$
by the affine structure (see
Lemma \ref{connection induced by a parallelisation})
coincides with the Riemannian connection induced by both metrics
$g\mul$
and
$G\mul \,.$
This is true although the two metrics need not to be proportional.
\subsubsection{Splittings of the configuration space}
\label{Splittings of the configuration space}
The multi-configuration space has two distinguished splittings.

\myskip

{\bf Multi--splitting.}
We have the obvious affine \emp{multi-splitting}
$\f P\mul = \f P_1 \car \dots \car \f P_n \,.$
The corresponding affine projections
$\pi_i : \f P\mul \to \f P_i$
and the further induced projections can be used to extract information
on the single particles from the kinematical and dynamical
multi--objects of the multi--system.

The subspaces
$\f S_1,\dots,\f S_n \sub \f S\mul$
are mutually orthogonal with respect to both metrics
$g\mul$
and
$G\mul \,.$

\myskip

{\bf Diagonal splitting.}
Moreover,
$\f P\mul$
has the following further \emp{diagonal splitting}, which
has no analogous in the one-particle scheme.

We define the \emp{diagonal affine subspace}, the
\emp{diagonal vector subspace} and the \emp{relative vector subspace},
respectively, as
\bat{2}
\f P\dia
&\byd
\{p\mul \in \f P\mul \mid p_1 = \dots = p_n \}
&&\sub
  \f P\mul \,,
\\
\f S\dia
&\byd
\{v\mul \in \f S\mul \mid v_1 = \dots = v_n\}
&&\sub
\f S\mul \,,
\\
\f S\rel
&\byd
\{v\mul \in \f S\mul \mid \sum_i\mu_i \, v_i = 0\}
&&\sub
\f S\mul \,.
\end{alignat*}

In the following, the subscripts
$``\dia"$
and
$``\rel"$
will denote the objects associated with the above spaces.

\bTh\label{diagonal splitting}
We have the affine splitting of the multi--configuration space and
the linear splittings of the associated vector and covector
multi--spaces
\bAt{4}
\f P\mul
&=
\f P\dia \drs \f S\rel
&&:
p\mul
&&=
(p_0, \dots, p_0) + (p_1-p_0, \, \dots, \, p_n-p_0)
\,,
\\
\f S\mul
&=
\f S\dia \drs \f S\rel
&&:
v\mul
&&=
(v_0, \dots, v_0) + (v_1-v_0, \, \dots, \, v_n-v_0)
\,,
\\
\f S^*\mul
&=
\f S^*\dia \drs \f S^*\rel
&&:
\alp\mul
&&=
(\mu_1 \, \alp_0, \, \dots, \, \mu_n \, \alp_0) +
(\alp_1-\mu_1 \, \alp_0, \, \dots, \, \alp_n-\mu_n \, \alp_0) \,,
\end{alignat}
where, for each
$p\mul \in \f P\mul \,,$
$p_0 \in \f P$
is the unique point such that
\[
\sum_i \mu_i \, (p_i - p_0) = 0 \,,
\]
and where, for each
$v\mul \in \f S\mul$
and
$\alp\mul \in \f S\mul^* \,,$
\[
v_0 \byd \sum_i \mu_i \, v_i
\ssep{and}
\alp_0 \byd \sum_i \alp_i \,.
\]

The above splittings are orthogonal with respect to the weighted
metric
$G\mul \,.$
\eTh

\bpf
1) Let us prove the splitting
$\f S\mul = \f S\dia \drs \f S\rel \,.$

For each
$v_0 \in \f S \,,$
we have
$\sum_i \mu_i v_0 = 0$
if and only if
$v_0 = 0 \,.$
Hence,
$\f S\dia \cap \f S\rel = 0 \,.$

Moreover, for each
$v\mul \in \f S\mul \,,$
we have
$(v_1, \dots, v_n) =
(v_0, \dots, v_0) + (v_1-v_0, \, \dots, \, v_n-v_0) \,,$
with any
$v_0 \in \f S \,.$
Clearly,
$(v_0, \dots, v_0) \in \f S\dia$
and
$(v_1-v_0 \, \dots, \, v_n-v_0) \in \f S\rel \,,$
if and only if
$v_0 = \sum_i \mu_i v_i \,.$
Hence,
$\f S\dia + \f S\rel = \f S\mul$
and the expression of the splitting is expressed by the 2nd formula
of the statement.

2) Let us prove the splitting
$\f P\mul = \f P\dia \drs \f S\rel \,.$

Clearly,
$\f P\dia \sub \f P\mul$
is an affine subspace associated with the vector subspace
$\f S\dia \sub \f S\mul \,.$
Hence, the equality
$\f S\mul = \f S\dia \drs \f S\rel$
implies
$\f P\mul = \f P\dia \drs \f S\rel \,.$

Moreover, for each
$p\mul \in \f P\mul \,,$
we have
$(p_1, \dots, p_n) = (p_0, \dots, p_0) + (p_1-p_0, \dots, p_n-p_0)\,,$
with any
$p_0 \in \f P \,.$
Clearly,
$(p_0, \dots, p_0) \in \f P\dia$
and
$(p_1-p_0, \dots, p_n-p_0) \in \f S\rel$
if and only if
$\sum_i \mu_i \, (p_i - p_0) = 0 \,.$
Hence, the expression of the splitting is expressed by the 1st formula
of the statement.

3) The splitting
$\f S\mul = \f S\dia \drs \f S\rel$
implies the splitting
$\f S^*\mul = \f S^*\dia \drs \f S^*\rel \,.$

Moreover, for each
$\alp\mul \in \f S\mul^* \,,$
we have
$(\alp_1, \dots, \alp_n) =
(\bet_1, \dots, \bet_n) + (\alp_1-\bet_1, \dots, \alp_n-\bet_n) \,,$
with any
$\bet_i \in \f S^*_i \,.$

On the other hand, for each
$v\mul \in \f S\mul \,,$
we obtain
\[
(\bet_1, \dots, \bet_n)(v_1, \dots, v_n) =
(\alp_1, \dots, \alp_n)(v_0, \dots, v_0)
\]
if and only if
\[
\sum_i \bet_i \, (v_i) =
\sum_j \alp_j \, (\sum_i \mu_i \, v_i) =
\sum_i \mu_i \, (\sum_j \alp_j) (v_i) \,,
\]
i.e., if and only if
\[
\bet_i = \mu_i \, (\sum_j \alp_j) \,.
\]

Moreover, in virtue of the equalities
\[
\sum_i \big(\alp_i - \mu_i \, (\sum_j \alp_j)\big) (v_i) =
\sum_i \alp_i \, (v_i) - \sum_j \alp_j \, (\sum_i \mu_i \, v_i) =
\sum_i \alp_i (v_i - v_0) \,,
\]
we obtain
\[
\big(
\alp_1 - \mu_1 \, (\sum_j \alp_j), \dots,
\alp_n - \mu_n \, (\sum_j \alp_j)\big) (v_1, \dots, v_n) =
(\alp_1, \dots, \alp_n)(v_1-v_0, \dots, v_n-v_0) \,.
\]

Hence, the expression of the splitting is expressed by the 3rd formula
of the statement.\QED
\epf

We denote the projection associated with the above splittings by
\bat{5}
\pi\dia
&:
\f P\mul
&&\to \f P\dia \,,
\qquad
&&\pi\rel
&&:
\f P\mul
&&\to \f S\rel \,,
\\
\ba\pi\dia
&:
\f S\mul
&&\to \f S\dia \,,
\qquad
&&\ba\pi\rel
&&:
\f S\mul
&&\to \f S\rel \,,
\\
\ba\pi^*\dia
&:
\f S^*\mul
&&\to \f S^*\dia \,,
\qquad
&&\ba\pi^*\rel
&&:
\f S^*\mul
&&\to \f S^*\rel \,.
\end{alignat*}

Clearly, the above splittings depend on the choice of the multi-mass
and are not orthogonal with respect to the geometrical metric
(unless all masses are equal).

We stress that, while we have the natural inclusion
$\f P\dia \hto \f P\mul \,,$
we do not have a natural inclusion
$\f S\rel \hto \f P\mul \,.$

We have a natural splitting of the weighted multi--metric of the type
$G\mul = G\dia \drs G\rel \,.$
Moreover, the affine structures of
$\f P\dia$
and
$\f S\rel$
yield the flat connections
$\nab\dia$
and
$\nab\rel \,,$
which turn out to be the Riemannian connections induced by
$G\dia$
and
$G\rel \,,$
respectively.

\myskip

{\bf Center of mass splitting.}
We can describe the diagonal splitting in another way, via the center
of mass.

According to the above Theorem, we define the \emp{center of mass} of
$p\mul \in \f P\mul$
to be the unique point
$p_0 \in \f P \,,$
such that
$\sum_i \mu_i \, (p_i - p_0) = 0 \,.$
By considering any
$o \in \f P \,,$
we can write
$p_0 = o + \sum_i \mu_i \, (p_i - o) \,.$
With reference to the center of mass, it is convenient to consider
a copy of the following pattern objects:
$\f P\cen \eqv \f P \,,\,$
$\f S\cen \eqv \f S$
and
$\f S^*\cen \eqv \f S^* \,.$

Thus, we have the \emp{center of mass} affine projection
\[
\pi\cen : \f P\mul \to \f P\cen :
p\mul \mto p_0 \byd o + \sum_i \mu_i \, (p_i - o) \,,
\qquad
\text{for any }
o \in\f P \,.
\]

The linear projections associated with the affine projection
$\pi\cen : \f P\mul \to \f P\cen$
turn out to be, respectively, the \emp{weighted sum} and the \emp{sum}
\bat{3}
\ba\pi\cen
&: \f S\mul \to \f S\cen
&&:
v\mul \mto v_0
&&\byd \sum_i \mu_i \, v_i \,,
\\
\E S\cen
&:
\f S\mul^* \to \f S\cen^*
&&:
\alp\mul \mto \alp_0
&&\byd \sum_i \alp_i \,.
\end{alignat*}

The 2nd projection is just the map which associates with every
multi--form its \emp{total value}.
Indeed, this map plays an important role in mechanics of
multi--systems.

Clearly, we have the natural affine isomorphism and linear
isomorphisms
\bat{3}
\f P\cen
&\to \f P\dia
&&: p_0
&&\mto (p_0, \dots, p_0) \,,
\\
\f S\cen
&\to \f S\dia
&&: v_0
&&\mto (v_0, \dots, v_0) \,,
\\
\f S^*\cen
&\to \f S^*\dia
&&: \bet_0
&&\mto
(\mu_1 \, \bet_0 \,,\, \dots \,,\, \mu_n \, \bet_0) \,.
\end{alignat*}

\bCr\label{center of mass splitting}
We have the \emp{center of mass splittings}
\bat{4}
\f P\mul
&\seq
\f P\cen \car \f S\rel
&&:
p\mul
&&\seq
\big(p_0 \,,\; (p_1-p_0,\, \dots,\, p_n-p_0)\big)
\,,
\\
\f S\mul
&\seq
\f S\cen \car \f S\rel
&&:
v\mul
&&\seq
\big(v_0 \,,\; (v_1-v_0, \, \dots, \, v_n-v_0)\big)
\,,
\\
\f S^*\mul
&\seq
\f S^*\cen \car \f S^*\rel
&&:
\alp\mul
&&\seq
\big(\alp_0 \,, \;
(\alp_1-\mu_1 \, \alp_0, \, \dots, \, \alp_n-\mu_n \, \alp_0)\big)
\,.\END
\end{alignat*}
\eCr

According to our scheme, we define the \emp{center of mass
kinetic energy} and the \emp{center of mass kinetic momentum} as
\bat{4}
\C K\cen
&: \B T^{-1} \ten T\f P\cen
&&\to
(\B T^{-2} \ten \B L^2 \ten \B M) \ten \Rn
&&:
v\cen
&&\mto \tfr12 m_0 \, g\cen(v\cen, v\cen) \,,
\\
\C P\cen
&: \B T^{-1} \ten T\f P\cen
&&\to
(\B T^{-1} \ten \B L^2 \ten \B M) \ten T^*\f P\cen
&&:
v\cen
&&\mto m_0 \, g\cen(v\cen) \,.
\end{alignat*}

Moreover, we obtain the equalities
\begin{gather*}
\C K\mul(v\mul) = \C K\cen(v_0) + \C K\mul(v\rel) \,,
\\
\begin{alignat*}{3}
\C K\cen \com T\pi\cen
&=
\C K\mul \com T\pi\dia
&&:
v\mul \mto \tfr12 m_0 \, g\cen (v_0, v_0)
&&=
\tfr12 m_0 \, \sum_i \mu_i \, g (v_0, v_0) \,,
\\
\C P\cen \com T\pi\cen
&=
\E S\cen \com \C P\mul \com T\pi\dia
&&:
v\mul \mto m_0 \, g\cen\Fla (v_0)
&&=
m_0 \, \sum_i \mu_i \, g\Fla (v_0) \,.
\end{alignat*}
\end{gather*}
\subsubsection{Kinematics}
\label{Kinematics}
For the kinematics of our system of $n$ particles, we follow the
viewpoint of a multi-particle moving in a multi-space
(in analogy with a one-particle moving in the standard space).
Moreover, we take into account the center of mass splitting.

\myskip

We define the \emp{multi-phase space} as
$J_1\f P\mul \byd
\f T \car (\B T^{-1} \ten T\f P\mul) \,.$

A \emp{multi-motion} is defined to be a map
$s\mul : \f T \to \f P\mul \,.$

Of course, the multi-motion can be regarded as the family of
motions of the system:
$s\mul = (s_1,\, \dots,\, s_n) \,.$
This holds also for the derived quantities, like the
\emp{multi-velocity}
$ds\mul : \f T \to \B T^{-1} \ten T\f P\mul$
and the \emp{multi-acceleration}
$\nab\mul ds\mul : \f T \to \B T^{-2} \ten T\f P\mul \,.$

We can relate the multi-motion to the splitting
of center of mass (Corollary \ref{center of mass splitting}) by the
equalities
\bgt
s\mul =
s\dia + s\rel \seq
(s\cen \,, s\rel) \,,
\qquad
ds\mul =
ds\dia + ds\rel \seq
(ds\cen \,,  ds\rel).
\\
\nab\mul ds\mul =
\nab\dia \, ds\dia + \nab\rel \, ds\rel \seq
(\nab\cen \, ds\cen \,, \nab\rel \, ds\rel) \,.
\end{gather*}
\subsubsection{Dynamics}
\label{Dynamics}
In analogy with the case of one--particle, we define a
\emp{multi-force} to be a map
\[
F\mul : J_1\f P\mul
\to (\B T^{-2} \ten \B L^2 \ten \B M) \ten T^* \f P\mul \,.
\]

Again, the multi-force can be regarded as the family of
forces acting on each particle
$F\mul = (F_1,\, \dots,\, F_n) \,.$
In general, each of the components is defined on the whole phase
space. In the particular case when each component
$F_i$
of the multi-force depends only the $i$-th phase space, the
multi-force is said to be \emp{without interaction}.

We say that a multi-force
$F\mul$
fulfills the \emp{Newton's 3rd principle} if, for each
$1 \leq i \leq n \,,$
\[
F_i = \sum_{1 \leq i \neq j \leq n}F_{ij} \,,
\qquad
F_{ij}(p_i, p_j) =
\lam_{ij}(\|p_j - p_i\|_{g}) \; g\Fla(p_j - p_i) \,,
\qquad
\lam_{ij} = \lam_{ji} \,,
\,,
\]
where
$F_{ij} : \f P_i \car \f P_j \to
(\B T^{-2} \ten \B L^2 \ten \B M) \ten T^* \f P_i$
and
$\lam_{ij} : \B L^2 \ten \Rn \to \Rn \,,$
for each
$1 \leq i \neq j \leq n \,.$

The \emp{total force} of the system is defined to be the
component of the multi-force with respect to the center of mass
\[
F\cen \byd \C S\cen \circ F\mul = \sum_i F_i
: J_1\f P\mul \to (\B T^{-2} \ten \B L^2 \ten \B M) \ten T^*\f P\cen
\,.
\]

The multi-force is said to be \emp{conservative} if it can be
derived from a \emp{multi-potential}
$\C U\mul : \f P\mul \to (\B T^{-2} \ten \B L^2 \ten \B M) \ten \Rn$
as
$F\mul  = d\C U\mul \,.$
In this case, we define the \emp{multi-Lagrangian} to be the map
\[
\C L\mul \byd \C K\mul + \C U\mul :
T\f P\mul \to (\B T^{-2} \ten \B L^2 \ten \B M) \ten \Rn \,.
\]

We say that a multi-motion
$s\mul$
fulfills the \emp{Newton's law of motion} if
\bEq
m_0 \, G\mul\Fla \com (\nab\mul \, ds\mul) = F\mul \com j_1s\mul \,.
\eEq

We can split the Newton's law with respect to the multi-splitting
and to the splitting of the center of mass.

In the former case, we simply obtain the system of coupled equations
\[
m_i \, g\Fla \com ds_i = F_i \com j_1s \,.
\]

In the latter case, we have the following Theorem.

\bTh
The Newton's equation is equivalent to the system
\bEq
m_0 \, g\cen\Fla \com (\nab\cen \, ds\cen) = F\cen \com j_1s\mul \,,
\qquad
m_0 \, G\rel\Fla \com (\nab\rel \, ds\rel) = F\rel \com j_1s\mul
\,.\END
\eEq
\eTh

If
$F\cen$
factors through
$J_1\f P\cen \,,$
then the 1st equation can be integrated independently of the 2nd
one and can be interpreted as the equation of motion of the center of
mass.

As for one-particle, the coordinate expression of the Newton's law
is, with reference to any chart
$(x^i\mul)$
of
$\f P\mul \,,$
\[
D (\dt\der_i \C K\mul \com ds\mul) - \der_i \C K\mul \com ds\mul
=  F_i \com j_1s\mul \,,
\]
which, if
$F\mul$
is conservative, is equivalent to the system of Lagrange's  equations
\[
D (\dt\der_i \C L\mul \com ds\mul) -\der_i \C L\mul \com ds\mul = 0
\,,
\]
\subsection{Constrained particles}
\label{Constrained particles}
\bsm
According to our programme, the analysis of the geometry of the
constrained space for a system of $n$  particles can be carried
out by analogy with the case of one-particle.
\esm

We assume the \emp{multi-configuration space} of a constrained
system of $n$ particles to be an embedded submanifold
$\f P\mul\,\cst \sub\f P\mul \,.$

In general, it is not possible to write
$\f P\mul\,\cst = \f P\cst\,_1 \car \dots \car \f P\cst\,_n \,,$
with
$\f P\cst\,_i \sub \f P_i \,,$
for each
$i = 1, \dots, n \,.$
In the particular case when this holds, we say that the constraint is
\emp{without interference between particles}.

Moreover, in general, it is not possible to write
$\f P\mul\,\cst = \f P\cen\,\cst \car \f S\rel\,\cst \,,$
with
$ \f P\cen\,\cst \sub \f P\cen$
and
$\f S\rel\,\cst \sub \f S\rel \,.$
In the particular case when this holds, we
say that the constraint is \emp{without interference between center
of mass and relative positions}.
In this case, the intrinsic metric
$G\cst$
splits into the sum of the metrics
$G\cen \, \cst$
and
$G\rel\,\cst$
(according to Corollary \ref{center of mass splitting}), with
interesting consequences in dynamics.

\myskip

We leave to the reader the task to formulate the kinematics and
dynamics of a constrained system of $n$ particles according to our
scheme.
\newpage
\section{Rigid systems}
\label{Rigid systems}
\bsm
Now, we specialise the theory of constrained systems of
$n$
particles to the case of a rigid constraint.
We devote emphasis to the geometric structure of the rigid
configuration space.
In particular, we show that this is the true source of the
classical formulas of the velocity of rigid systems.

Throughout the section, we suppose that
$n\geq 2 \,.$
\esm
\subsection{Geometry of the configuration space}
\label{Geometry of the configuration space}
Let us define the scaled functions, for $1\leq i, j\leq n \,,$
\[
r_{ij} : \f P\mul \to \B L \ten \Rn :
p\mul \eqv \npd{p}{n} \mto \| p_i - p_j\|_{g}\,.
\]

A \emp{rigid configuration space\/} is defined to be a
subset of the type
\[
i\rig : \f P\rig \byd
\{p\mul \in \f P\mul \sst r_{ij}(p\mul) = l_{ij} \,,
1 \leq i< j \leq n \} \sub \f P\mul \,,
\]
where
$l_{ij} \in \B L$
fulfill
\bat{3}
l_{ij}
&= l_{ji}\,,
&&\qquad
1\leq i, j\leq n \,,
&&\quad
i\neq j\,,
\\
l_{ik}
&\leq l_{ij} + l_{jk}\,,
&&\qquad
1\leq i,j,k\leq n \,,
&&\quad
i\neq j,j \neq k,k \neq i\,.
\end{alignat*}

Note that we have excluded the case in which the positions of
different particles coincide.

 From now on, let us consider a given rigid configuration space
$\f P\rig \,.$

\myskip

We define the \emp{rotational space} to be the subset
\[
i\rot : \f S\rot \byd
\{v\rel \in \f S\rel \sst \| v_i - v_j \| = l_{ij} \,,
1 \leq i < j \leq n \} \sub \f S\rel \,.
\]

A typical notation for the elements of
$\f S\rot$
will be
\[
r\rot = (r_1, \dots, r_n) \in \f S\rot \sub \f S\rel \,.
\]

Due to the equality
$\|p_i - p_j\| =
\|\pi\rel(p\mul)_i - \pi\rel(p\mul)_j\| = l_{ij} \,,$
the rigid constraint does not involve the center of mass but only
relative positions.
Then, the restrictions of the projections
$\pi\cen : \f P\mul \to \f P\cen$
and
$\pi\rel : \f P\mul \to \f S\rel$
to the subset
$\f P\rig \sub \f P\mul$
turn out to be, respectively, projections
\[
\pi\cen : \f P\rig \to \f P\cen
\ssep{and}
\pi\rot : \f P\rig \to \f S\rot \,.
\]

Thus, we obtain the bijection
\[
(\pi\cen \,,\,\pi\rot) :
\f P\rig \to \f P\cen \car \f S\rot \,.
\]

Next, we classify the rigid constraints as follows.

For each
$v\rel \in \f S\rel \,,$
we define the vector subspace
\[
\la v\rel \ra \byd
\Span \{v_i - v_j \st 1 \leq i,j \leq n\} \sub \f S \,.
\]

\bLm
For each
$v\rel \in \f S\rel \,,$
we have
$\la v\rel \ra =
\Span \{ v_i-v_h \mid h = 1, \dots, n\} \,,$
for any chosen
$1 \leq i \leq n \,.$
More precisely,
$v_i = \sum_{h \neq i} \, \mu_h \, (v_i - v_h) \,.$
\eLm

\bpf
We have
$- \mu_i v_i = \sum_{h \neq i}\, \mu_h v_h$
and
$v_i - \mu_i v_i = \sum_{h \neq i}\, \mu_h v_i \,,$
which give the result by subtraction.\QED
\epf

If
$r\rot \in \f S\rot \,,$
then we call
$\lra {r\rot}$
the \emp{characteristic space} of
$\lra {r\rot} \,.$

Let us prove that the dimension of the characteristic spaces
does not depend on elements in
$\f S\rot \,,$
but only on
$\f S\rot \,.$
We need the following technical Lemma (see also \cite{CurMil85}).

\bLm\label{isometry linking to rotational configurations}
Let
$r\rot \,,\, r\rot' \in \f S\rot \,.$
Then, there is an isometry
$\phi :\f S \to \f S$
such that
$\phi(r_i) = r'_i \,,$
for each
$i = 1, \dots, n \,.$
\eLm

\bpf
Let us fix
$i \in \{1, \dots, n\} \,,$
and set
$\dim \lra {r\rot} = l \,.$
We can choose a basis of
$\la v\rot \ra$
of the type
$(b_h) \byd \{r_i - r_{i_h} \mid h = 1, \dots, l\} \,.$
Consider the subset
$(b'_h) \byd \{r'_i - r'_{i_h} \mid h = 1, \dots, l\} \,.$
We can define a linear map
$\phi : \lra {r\rot} \to \lra {r\rot'}$
by
$\phi(b_i) = b'_i \,,$
for each
$h = 1, \dots, l \,.$
We have
$g(\phi(b_i), \phi(b_j)) = g(b_i, b_j) \,,$
in virtue of the equalities
$\|b_h \| = l_{ih} = \| b'_h \|$
and
$\| b_h - b_k\| = l_{hk} = \| b'_h - b'_k\| \,.$
The map
$\phi$
is an isometry between
$\lra {r\rot}$
and
$\lra {r\rot'} \,,$
hence it can be extended to a linear isometry of
$\f S \,.$
Eventually, we have
$g(\phi(r_j), \phi(b_h)) = g(r_j, b_h) \,,$
and the rigid constraint yields
$g(r_j, b_h) = g(r'_j, \phi(b_h)) \,.$
Hence, we obtain
$\phi(r_j) = r'_j \,,$
for each
$j = 1, \dots, n \,.$\QED
\epf

Therefore, if
$r\rot \,, r\rot' \in \f S\rot \,,$
then
$\dim \lra {r\rot} = \dim \lra {r\rot'} \,.$

We define the \emp{characteristic\/} of
$\f P\rig$
to be the integer number
$C_{\f P\rig} \byd \dim \lra {r\rot} \,,$
where
$r\rot \in \f S\rot \,.$
Obviously, we have
$1 \leq C_{\f P\rig} \leq 3 \,,$
and we can classify the rigid configuration space in terms of
$C_{\f P\rig} \,.$

We say
$\f P\rig$
to be

-- \emp{strongly non degenerate\/} if
$C_{\f P\rig} = 3 \,,$

-- \emp{weakly non degenerate\/} if
$C_{\f P\rig} = 2 \,,$

-- \emp{degenerate\/} if
$C_{\f P\rig} = 1 \,.$

Of course, if
$n = 2 \,,$
then
$\f P\rig$
is degenerate; if
$n = 3 \,,$
then
$\f P\rig$
can be degenerate or weakly non degenerate.

Thus, by considering all particles as assuming positions in the
same space $\f P \,,$ the above cases correspond respectively to
the case when the minimal affine subspace containing all particles is
a line, or a plane, or the whole
$\f P \,.$
As a consequence of the above result, the case occurring for a given
rigid system does not change during the motion.

\myskip

Let us denote by
$O(\f S,g)$
the group of orthogonal transformations of
$\f S$
with respect to
$g \,.$
We want to study the topological subspace
$\f S\rot \sub \f S\rel$
through the natural action of the Lie group
$O(\f S,g)$
on
$\f S\rot \,.$
More precisely, we can easily prove that the map
\[
O(\f S,g) \car \f S\rot \to \f S\rot : (\phi,r\rot)
\mto \big(\phi(r_1), \dots, \phi(r_n)\big)
\]
is well--defined and yields a continuous action of
$O(\f S,g)$
on
$\f S\rot \,.$
Such an action of
$O(\f S,g)$
on
$\f S\rot$
is transitive because of
Lemma \ref{isometry linking to rotational configurations}.
Let us denote the isotropy group at
$r\rot$
by
$H(r\rot) \sub O(\f S,g) \,.$

\bLm
The following facts hold:

- in the {\em strongly non degenerate case\/} the isotropy subgroup
$H[r\rot]$
is the trivial subgroup
$\{1\} \,;$

- in the {\em weakly non degenerate case\/} the isotropy subgroup
$H[r\rot]$
is the discrete subgroup of reflections with respect to
$\lra {r\rot} \,;$

- in the {\em degenerate case\/} the isotropy subgroup
$H[r\rot]$
is the 1 dimensional subgroup of rotations whose axis is
$\lra {r\rot} \,;$
we stress that this subgroup is not normal.\END
\eLm

\bPr
The following facts hold:

-- $\f S\rot$
is {\em strongly non degenerate\/} if and only if the action of
$O(\f S, g)$
on
$\f S\rot$
is free;

-- $\f S\rot$
is {\em weakly non degenerate\/} if and only if the action of
$O(\f S, g)$
on
$\f S\rot$
is not free, but the action of the subgroup
$SO(\f S, g) \sub O(\f S, g)$
on
$\f S\rot$
is free;

-- $\f S\rot$
is {\em degenerate\/} if and only if the action of
$SO(\f S, g)$
on
$\f S\rot$
is not free.\END
\ePr

Of course, if
$n = 2 \,,$
then
$\f S\rot$
is degenerate; if
$n = 3 \,,$
then
$\f S\rot$
can be degenerate or weakly non degenerate.

Furthermore, we can prove the following result (recall the definition
of affine space in Section \ref{Generalised affine spaces}).

\bCr
The following facts hold:

-- if
$\f S\rot$
is {\em strongly non degenerate\/}, then
$\f S\rot$
is an affine space associated with the group
$O(\f S, g) \,;$

-- if
$\f S\rot$
is {\em weakly non degenerate\/}, then
$\f S\rot$
is an affine space associated with the group
$SO(\f S, g) \,;$

-- if
$\f S\rot$
is {\em degenerate\/}, then
$\f S\rot$
is a homogeneous space (i.e. the quotient of a Lie group with respect
to a closed subgroup) with two possible distinguished diffeomorphisms
(depending on a chosen orientation on the straight line of the rigid
system) with the unit sphere
$\fE S^2 \sub \B L^* \ten \f S \,,$
with respect to the metric
$g \,.$

Thus, in all cases
$\f S\rot$
turns out to be a manifold (see
Section \ref{Generalised affine spaces}
and
\cite{War71}).\END
\eCr

\bCr
In the non degenerate case, the choice of a configuration
$r\rot \in \f S\rot$
and of a scaled orthonormal basis in
$\B L^* \ten \f S$
yield the diffeomorphisms (via the action of
$O(\f S, g)$
on
$\f S\rot$)
\bat{3}
\f S\rot
&\seq
O(\f S, g)
&&\seq
O(3) \,,
&&\quad
\text{in the strongly non degenerate case;}
\\
\f S\rot
&\seq
SO(\f S, g)
&&\seq
SO(3) \,,
&&\quad
\text{in the weakly non degenerate case}.
\end{alignat*}

In the degenerate case, the continuous choice of an orientation on
the straight lines
$\lra {r\rot} \sub \f S$
generated by each configuration
$r\rot \in \f S\rot$
and of a scaled orthonormal basis in
$\B L^* \ten \f S$
yields the diffeomorphisms
\[
\f S\rot \seq \fE S^2(\B L^* \ten \f S, g) \seq \fE S^2(3) \,,
\quad
\text{in the degenerate case.}\END
\]
\eCr

 From now on, in the non degenerate case, we shall refer only to
one of the two connected components of
$\f S\rot \,,$
for the sake of simplicity and for physical reasons of continuity.
Accordingly, we shall just refer to the non degenerate case (without
specification of strongly or weakly non degenerate), or to the
degenerate case.
\subsection{Tangent space of rotational space}
\label{Tangent space of rotational space}
\bsm
The rotational space
$\f S\rot$
is embedded into the environmental relative vector space
$\f S\rel \,.$
Hence, the tangent vectors of
$\f S\rot$
can be regarded as multi--vectors of
$\f S\rel \,,$
which respect the rigid constraint.
Here, we show a very geometric way to describe the tangent vectors of
$\f S\rot \,,$
so obtaining a geometric interpretation of classical formulas of the
mechanics of rigid systems.
\esm
\subsubsection{Non degenerate case}
\label{Non degenerate case}
First, we recall the following well--known result (see, for instance,
\cite{War71}).

\bLm\label{orthogonal Lie algebra}
By regarding the orthogonal group as a subspace
$O(\f S, g) \sub L(\f S, \f S) \,,$
we can identify the Lie algebra
$so(\f S, g)$
with the subspace
$so(\f S, g) \sub \f S^* \ten \f S \,,$
consisting of the tensors which are antisymmetric with respect to
$g \,.$
\eLm

\bpf
It is well--known that the subspace
$O(\f S, g) \sub L(\f S, \f S)$
consists of the invertible elements
$f \in L(\f S, \f S) \,,$
which are invertible and fulfill the condition
$f^{-1} = f^\E t \,.$

Hence, if
$c : \Rn \to O(\f S, g) \sub L(\f S, \f S)$
is a curve such that
$c(0) = \id \,,$
then we obtain
$\id = c \com c^{-1} = c \com c^\E t \,,$
hence
$0 = D(c \com c^\E t)(0) = Dc(0) + Dc^\E t(0) \,. $
Thus, the vectors tangent to
$O(\f S, g)$
at the identity consist of antisymmetric endomorphisms.

Conversely, we can prove that, for each antisymmetric
endomorphism
$\ome \in L(\f S, \f S) \,,$
there is a curve
$c : \Rn \to O(\f S, g) \sub L(\f S, \f S) \,,$
such that
$c(0) = \id$
and
$Dc(0) = \ome \,.$\QED
\epf

\bPr\label{rotational parallelisation}
We have the natural parallelising linear isomorphism
\bEq
\tau\rot : T\f S\rot \to \f S\rot \car so(\f S, g)
\eEq
and the associated projection
\[
\rho\rot : T\f S\rot \sub \f S\rot \car \f S\rel \to so(\f S, g) :
(r\rot \,, \; v\rot) \mto \ome \,.
\]

The expression of the inverse isomorphism
$\tau\rot^{-1}$
is
\[
\tau\rot^{-1}
: \f S\rot \car so(\f S, g) \to T\f S\rot \sub \f S\rot \car \f S\rel
: (r\rot, \, \ome) \mto
\Big(r\rot, \, \big(\ome(r_1), \dots,\ome(r_n)\big)\Big) \,.
\]

Thus, for each
$(r\rot, v\rot) \in T\f S\rot
\sub
\f S\rot \car \f S\rel \,,$
there is a unique
$\ome \in so(\f S, g) \,,$
such that
$v_i = \ome(r_i) \,,$
for
$1 \leq i \leq n \,.$
\ePr

\bpf
By recalling that
$\f S\rot$
is an affine space associated with the Lie group
$O(\f S, g) \,,$
Lemma \ref{parallelisation of an affine space}
yields the isomorphisms
$\tau\rot \,.$

Next, in order to compute
$\tau\rot^{-1} \,,$
let us consider an element
$\ome \in so(\f S, g) \,.$
Then, there exists a map
$\ti\ome : \Rn \to SO(\f S, g)$
such that
$\ti\ome(0) = \id \in SO(\f S, g)$
and
$(D\ti\ome)(0) = \ome \,.$
Hence, for each
$r\rot \in \f S\rot \sub \f S\rel \,,$
the curve
$c : \Rn \to \f S\rel :
\lam \mto \big(\ti\ome(\lam)(r_1), \dots, \ti\ome(\lam)(r_n)\big)$
is valued in
$\f S\rot$
because
\[
\|\ti\ome(\lam)(r_i) - \ti\ome(\lam)(r_j)\| =
\|\ti\ome(\lam)(r_i - r_j)\| =
\|r_i - r_j\| \,,
\qquad
\Al \lam \in \Rn \,,
\quad
\Al 1 \leq i, j \leq n \,.
\]
Hence, the tangent map
$Dc(0) \in \f S\rel$
is valued in
$T_{r\rot}\f S\rot \in \f S\rel \,.$
On the other hand, we have
\[
Dc(0) = \big(\ome(r_1), \dots, \ome(r_n)\big) \,.\QED
\]
\epf

Later,
we shall give an explicit expression of the parallelisation
$\tau\rot \,,$
via the ``inertia isomorphism"
(Corollary \ref{expression of rotational parallelisation}).

\myskip

We can read the parallelisation
$\tau\rot$
in a further interesting way, by means of an algebraic
re--interpretation of
$so(\f S, g) \,.$

For this purpose, we recall the cross products
$\cro$
of
$\f S$
and of
$\f S^* \,,$
defined by
\bat{4}
u \cro v
&\byd *(v \wed w)
&&\byd
g\Sha (i(u \wed v) \, \eta)
&&= i\big(g\Fla(u) \wed g\Fla(v)\big) \, \ba\eta \,,
\qquad
&&\Al \, u, v \in \f S \,,
\\
\alp \cro \bet
&\byd *(\alp \wed \bet)
&&\byd
g\Fla (i(\alp \wed \bet) \, \ba\eta)
&&= i\big(g\Sha(\alp) \wed g\Sha(\bet)\big) \, \eta \,,
\qquad
&&\Al \, \alp, \bet \in \f S^* \,.
\end{alignat*}

The cross product commutes with the metric isomorphisms, i.e. we have
\[
g(u \cro v) =
g\Fla(u) \cro g\Fla(v)
\ssep{and}
g\Sha(\alp \cro \bet) =
g\Sha(\alp) \cro g\Sha(\bet) \,.
\]

For short, we set
\bat{3}
u \cro\mul v\mul
&\byd (u \cro v_1, \dots, u \cro v_n) \,,
&&\qquad
\Al \, u \in \f S \,,
\quad
&&\Al \, v\mul \in \f S\mul \,,
\\
\alp \cro\mul \bet\mul
&\byd (\alp \cro \bet_1, \dots, \alp \cro \bet_n) \,,
&&\qquad
\Al \, \alp \in \f S^* \,,
\quad
&&\Al \, \bet\mul \in \f S^*\mul \,.
\end{alignat*}

Moreover, we introduce the scaled vector space
\bEq
\f V\anl \byd \B L^{-1} \ten \f S \,.
\eEq

\bLm\label{Hodge isomorphism}
The metric isomorphism
$g\Fla : \f S^* \ten \f S \to \B L^{2} \ten \f S^* \ten \f S^* :
\alp \ten v \mto \alp \ten g(v, \cdot)$
and the Hodge isomorphism
$* : \B L^{2} \ten \Lam^2 \f S^* \to \B L^{-1} \ten \f S :
\ome \mto i(\ome) \, \ba\eta$
yield the linear isomorphism
\[
* \com g\Fla : so(\f S, g) \to \f V\anl : \ome \mto
i\big(g\Fla(\ome)\big) \ba\eta \,.\END
\]
\eLm

\bCr\label{angular parallelisation}
We have the natural parallelising isomorphism
\bEq
\tau\anl \byd * \com g\Fla \com \tau\rot :
T\f S\rot \to \f S\rot \car \f V\anl
\eEq
and the associated projection
\[
\rho\anl : T\f S\rot \sub \f S\rot \car \f S\rel \to \f V\anl :
(r\rot, \; v\rot) \mto \Ome \,.
\]

The expression of the inverse isomorphism
$\tau\anl^{-1}$
is
\[
\tau\anl^{-1} : \f S\rot \car \f V\anl \to T\f S\rot
\sub \f S\rot \car \f S\rel :
(r\rot \,,\, \Ome) \mto
\big(r\rot \,, \, (\Ome \cro r_1, \dots, \Ome \cro r_n)\big) \,.
\]

Thus, for each
$(r\rot, v\rot) \in T\f S\rot
\sub
\f S\rot \car \f S\rel \,,$
there is a unique
$\Ome \in \f V\anl \,,$
such that
$v_i = \Ome \cro r_i \,,$
for
$1 \leq i \leq n \,.$
\eCr

\bpf
It follows from
Proposition \ref{rotational parallelisation}
and the
Lemma \ref{Hodge isomorphism}.\QED
\epf

Later, we shall give an explicit expression of the parallelisation
$\tau\anl \,,$
via the ``inertia isomorphism"
(Theorem \ref{expression of angular parallelisation}).

The above
Corollary \ref{angular parallelisation}
is just a geometric formulation of the well--known formula expressing
the relative velocity of the particles of a rigid system through the
angular velocity.

\bCr
The transpose
$(\tau\anl^{-1})^*$
of the isomorphism
$\tau\anl^{-1}$
has the expression
\bEq
(\tau\anl^{-1})^* : T^*\f S\rot \to \f S\rot \car \f V\anl^*:
(r\rot,\alp) \mto \big(r\rot, \; \sum_i g\Fla(r_i) \cro \alp_i\big)
\,.
\eEq
\eCr

\bpf
The expression of
$\tau\anl^{-1}$
and cyclic permutations yield
\begin{gather*}
(\tau\anl^{-1})^*(r\rot, \alp) \, (\Ome) \byd
\\
\begin{align*}
&=
\alp \big(\tau\anl(r\rot, \Ome)\big) =
\alp \, (\Ome \cro\mul r\rot) =
\sum_i \alp_i \, (\Ome \cro r_i) =
\sum_i g \, \big(g\Sha(\alp_i) \,, (\Ome \cro r_i)\big)
\\
&=
\sum_i g \, \big(r_i \,,\, (g\Sha_i(\alp_i) \cro \Ome)\big) =
\sum_i g \, \big(\Ome \,,\, (r_i \cro g\Sha_i(\alp_i))\big) =
\sum_i \big((g\Fla_i(r_i) \cro \alp_i\big) \, (\Ome) \,.\QED
\end{align*}
\end{gather*}
\epf

The cross product
$\cro$
is equivariant with respect to the action of
$SO(\f S, g) \,.$
Hence, the isomorphism
$\tau\anl$
turns out to be equivariant with respect to this group.
\subsubsection{Degenerate case}
\label{Degenerate case}
Let us consider the quotient vector bundle
$[\f S\rot \car so(\f S, g)]$
over
$\f S\rot$
of the vector bundle
$\f S\rot \car so(\f S, g) \,,$
with respect to the vector subbundle
$h[\f S\rot]$
consisting, for each
$r\rot \in \f S\rot \,,$
of the isotropy Lie subalgebra
$h[r\rot] \sub so(\f S,g)$
of
$\lra {r\rot} \sub \f S$
(see, for instance, \cite{War71}).

\myskip

Now, let us refer to the degenerate case.
We can rephrase the results concerning the non degenerate case by a
quotient procedure. In particular, we have the following results.

\myskip

For each
$r\rot \in \f S\rot \,,$
the isotropy Lie subalgebra associated with
$r\rot$
consists of antisymmetric endomorphisms
$\phi \in \f S^* \ten \f S$
which preserve the 1--dimensional vector subspace
$\lra {r\rot} \sub \f S$
generated by
$r\rot \,.$

\bLm
We have the natural linear fibred isomorphism
\[
[\tau\rot] : T\f S\rot \to [\f S\rot \car so(\f S, g)] \,.
\]
\eLm

\bpf
It follows from a well--known result on homogeneous spaces (see, for
instance, \cite{War71}).\QED
\epf

Let us consider the quotient vector bundle
$[\f S\rot \car \f V\anl]$
over
$\f S\rot$
of the vector bundle
$\f S\rot \car \f V\anl$
with respect to the vector subbundle
$a[\f S\rot]$
consisting, for each
$r\rot \in \f S\rot \,,$
by the 1--dimensional vector subspace
$\lra {r\rot} \sub \f V\anl$
generated by
$r\rot \,.$

\bPr
We have the linear fibred isomorphism
\[
[\tau\anl] : T\f S\rot \to [\f S\rot \car \f V\anl] \,.
\]

The expression of the inverse isomorphism
$[\tau\anl^{-1}]$
is
\bml
[\tau\anl]^{-1} :
[\f S\rot \car \f V\anl]
\to T\f S\rot \sub \f S\rot \car \f S\mul :
\\
: (r\rot \,,\, [r\rot,\Ome]) \mto
\big(r\rot \,,\, (\Ome \cro r_1, \dots, \Ome \cro r_n)\big)
\,,
\end{multline*}
where the cross products
$\Ome \cro r_i$
turn out to be independent of the choice of representative for the
class
$[r\rot, \Ome] \,.$

Thus, for each
$(r\rot, v\rot) \in T\f S\rot \sub \f S\rot \car \f S\rel \,,$
there is a unique
$[r\rot, \Ome] \in [\f S\rot \car \f V\anl]_{r\rot} \,,$
such that
$v_i = \Ome \cro r_i \,,$
for
$1 \leq i \leq n \,.$\END
\ePr

\bPr
A continuous choice of an orientation of the straight lines
$\lra {r\rot} \sub \f S$
generated by the configurations
$r\rot \in \f S\rot$
yields the linear isomorphism
\[
T\f S\rot \seq T\fE S^2(\B L^* \ten \f S, g) \,.\END
\]
\ePr
\subsection{Rigid system metrics}
\label{Rigid system metrics}
\bsm
The multi--dynamical metric of
$\f S\mul$
induces a metric on
$\f S\rot \,,$
which can be regarded also in another useful way through the
isomorphism
$\tau\anl \,,$
and will be interpreted as the inertia tensor.

Moreover, the standard pattern metric of
$\f V\anl$
induces a further metric on
$\f S\rot \,.$
\esm

\bPr
The inclusion
$i\rig : \f P\rig \hto \f P\mul$
yields the geometrical and weighted scaled Riemannian metrics
\bat{3}
g\rig
&\byd i^*\rig \, g\mul
&&: \f P\rig
&&\to \B L^2 \ten (T^*\f P\rig \ten T^*\f P\rig) \,,
\\
G\rig
&\byd i^*\rig \, G\mul
&&: \f P\rig
&&\to \B L^2 \ten (T^*\f P\rig \ten T^*\f P\rig) \,.
\end{alignat*}

The splitting
$\f P\rig = \f P\dia \drs \f S\rot$
is orthogonal with respect to the metric
$G\rig \,.$
\ePr

\bpf
The splitting
$\f P\mul = \f P\dia \drs \f S\rel$
is orthogonal with respect to the metric
$G\mul$
and we have
$\f P\rig = \f P\dia \drs \f S\rot \,,$
with
$\f S\rot \sub \f S\rel \,.$\QED
\epf

\bPr
The inclusion
$i\rot : \f S\rot \hto \f S\rel$
yields the geometrical and weighted scaled Riemannian metrics
\bat{3}
g\rot
&\byd i^*\rot \, g\rel
&&: \f S\rot
&&\to \B L^2 \ten (T^*\f S\rot \ten T^*\f S\rot) \,,
\\
G\rot
&\byd i^*\rot \, G\rel
&&: \f S\rot
&&\to \B L^2 \ten (T^*\f S\rot \ten T^*\f S\rot) \,.
\end{alignat*}

For each
$(\Ome \cro\mul r\rot) \,, \; (\Ome' \cro\mul r\rot)
\in T_{r\rot}\f S\rot \sub \f S\rel \,,$
we have the expressions
\bal
g\rot (r\rot) \, (\Ome \cro\mul r\rot \,,\, \Ome' \cro\mul r\rot)
&=
\sum_i \big(
g(r_i, r_i) \, g(\Ome, \Ome') -
g(r_i, \Ome) \, g(r_i, \Ome')\big) \,,
\\
G\rot (r\rot) \, (\Ome \cro\mul r\rot \,,\, \Ome' \cro\mul r\rot)
&=
\sum_i \mu_i \big(
g(r_i, r_i) \, g(\Ome, \Ome') -
g(r_i, \Ome) \, g(r_i, \Ome')\big) \,.
\end{align*}
\ePr

\bpf
In virtue of standard properties of the cross product, we obtain
\bal
g\rot (r\rot) \, (\Ome \cro\mul r\rot \,,\, \Ome' \cro\mul r\rot)
&\byd
\sum_i g(\Ome \cro r_i,\, \Ome' \cro r_i) =
\sum_i g\big((\Ome' \cro r_i) \cro \Ome \,, \; r_i\big)
\\
&=
\sum_i \big(
g(r_i, r_i) \, g(\Ome, \Ome') -
g(r_i, \Ome) \, g(r_i, \Ome')\big) \,,
\\
G\rot (r\rot) \, (\Ome \cro\mul r\rot \,,\, \Ome' \cro\mul r\rot)
&\byd
\sum_i \mu_i g(\Ome \cro r_i,\, \Ome' \cro r_i) =
\sum_i \mu_i g\big((\Ome' \cro r_i) \cro \Ome \,, \; r_i\big)
\\
&=
\sum_i \mu_i \big(
g(r_i, r_i) \, g(\Ome, \Ome') -
g(r_i, \Ome) \, g(r_i, \Ome')\big) \,.\QED
\end{align*}
\epf

We can regard the metrics
$g\rot$
and
$G\rot$
in another interesting way, via
$\tau\anl \,.$

\bCr\label{inertia metrics}
In the non degenerate case, the isomorphism
$\tau\anl$
allows us to read
$g\rot$
and
$G\rot$
as the scaled metrics
\bAt{2}
\sig
&\byd
(\tau\anl^{-1})^* \, g\rot :
\f S\rot \to \B L^2 \ten (\f V^*\anl \ten \f V^*\anl) \,,
\\
\Sig
&\byd
(\tau\anl^{-1})^* \, G\rot :
\f S\rot \to \B L^2 \ten (\f V^*\anl \ten \f V^*\anl) \,,
\end{alignat}
with expressions
\bat{2}
\sig
&(r\rot) (\Ome, \Ome')
&&=
\sum_i \big(
g(r_i,  r_i) \, g(\Ome,\Ome') -
g(r_i, \Ome) \, g(r_i, \Ome')
\big) \,,
\\
\Sig
&(r\rot) (\Ome, \Ome')
&&=
\sum_i \mu_i \, \big(
g(r_i, r_i) \, g(\Ome,\Ome') -
g(r_i, \Ome) \, g(r_i, \Ome')
\big) \,.\END
\end{alignat*}
\eCr

We have a further natural metric of
$\f S\rot \,.$
For this purpose, let us consider the metric
$g \in \f V^*\anl \ten \f V^*\anl$
of
$\f V\anl$
naturally induced by the pattern metric
$g$
of
$\f S \,.$
We can make the natural identification
$O(\f V\anl, g) \seq O(\f S, g) \,.$

\bPr
In the non degenerate case, we obtain the unscaled Riemannian metric
\[
g\anl \byd \tau\anl^* \, g :
\f S\rot \to T^*\f S\rot \ten T^*\f S\rot \,.
\]

For each
$(r\rot \,, \; \Ome \cro\mul r\rot) \,, \;
(r\rot \,,\: \Ome' \cro\mul r\rot)
\in T\f S\rot \sub \f S\rot \car \f S\rel \,,$
we have the expression
\[
g\anl (r\rot) \, (\Ome \cro\mul r\rot \,,\, \Ome' \cro\mul r\rot) =
g (\Ome, \Ome') \,.\END
\]
\ePr

\myskip

All metrics of
$\f S\rot$
considered above are invariant with respect to the left action of
$O(\f S, g) \,.$

\bPr\label{metric isomorphisms in the non degenerate case}
In the non degenerate case, the choice of a configuration
$r\rot \in \f S\rot$
and of an orthonormal basis in
$\f V\anl \,,$
respectively, yields the  following diffeomorphisms (via the action
of
$SO(\f V\anl, g)$
on
$\f S\rot$)
\[
\f S\rot \seq SO(\f V\anl, g) \seq SO(3) \,,
\]
which turn out to be isometries with respect to the Riemannian metrics
$g\anl \,,\, - \tfr12 k\anl$
and
$- \tfr12 k_3 \,,$
of
$\f S\rot \,,\, \f V\anl$
and
$SO(3) \,,$
where
$k\anl$
and
$k_3$
are the Killing forms.
\ePr

\bpf
The above diffeomorphisms yield the linear fibred isomorphisms
\[
T_{r\rot}\f S\rot \seq so(\f V\anl, g) \seq so(3) \,.
\]

On the other hand, the natural isomorphism
$so(\f V\anl, g) \to \f V\anl \,,$
induced by
$g\Fla$
and
$* \,,$
is metric.
Hence, in virtue of the definition of
$g\anl \,,$
the isomorphism
$T_{r\rot}\f S\rot \seq so(\f V\anl, g)$
turns out to be metric.

Moreover, the metric
$g$
of
$\f V\anl$
turns out to coincide with the metric
$- \tfr12 k\anl$
of
$so(\f V\anl, g) \,.$
In fact, we have
$g(\ome, \ome') =
- \tfr12 \, \tr\big((\Ome \cro) \com (\Ome' \cro)\big) \,.$
It is easy to see that the isomorphism
$so(\f V\anl, g) \seq so(3)$
is metric.\QED
\epf

\myskip

We leave to the reader the task to extend the above results to the
degenerate case.
\subsubsection{Angular automorphisms}
\label{Angular automorphisms}
\bPr
The unscaled metric
$g$
of
$\f V\anl$
allows us to regard the metrics
$\sig$
and
$\Sig$
as scaled symmetric fibred endomorphisms
\bat{2}
\ha\sig
&\byd
g\Sha \com \sig\Fla
&&:
\f S\rot \to \B L^2 \ten (\f V^*\anl \ten \f V\anl) \,,
\\
\ha\Sig
&\byd
g\Sha \com \Sig\Fla
&&:
\f S\rot \to \B L^2 \ten (\f V^*\anl \ten \f V\anl) \,.
\end{alignat*}

We have the expressions
\bat{2}
\ha\sig (r\rot)(\Ome)
&= \sum_i r_i \cro (\Ome \cro r_i)
&&=
\sum_i \big(g(r_i,r_i) \, \Ome - g(r_i,\Ome) \, r_i\big)
\\
\ha\Sig (r\rot)(\Ome)
&= \sum_i \mu_i \, r_i \cro (\Ome \cro r_i)
&&=
\sum_i \mu_i \, \big(g(r_i,r_i) \, \Ome - g(r_i,\Ome) \, r_i\big)
\,.
\end{alignat*}

In the non degenerate case, the above maps are automorphisms.
\ePr

\bpf
It follows immediately from
Corollary \ref{inertia metrics}.\QED
\epf

\bLm
In the non degenerate case, we have
\bgt
\ha\sig^{-1} = \sig\Sha \com g\Fla
\ssep{and}
\ha\Sig^{-1} = \Sig\Sha \com g\Fla \,,
\\
(\ha\sig^{-1})^* = g\Fla \com \sig\Sha
\ssep{and}
(\ha\Sig^{-1})^* = g\Fla \com \Sig\Sha \,.
\end{gather*}
\eLm

\bpf
We have
$\ha\sig^{-1} \byd
(g\Sha \com \sig\Fla)^{-1} =
(\sig\Fla)^{-1} \com (g\Sha)^{-1} =
\sig\Sha \com g\Fla \,.$

Moreover, the symmetry of
$g$
and
$\sig$
give
$(\ha\sig^{-1})^* =
(\sig\Sha \com g\Fla)^* = (g\Fla)^* \com (\sig\Sha)^* =
g\Fla \com \sig \Sha \,.$

Analogous proof holds for
$\Sig \,.$\QED
\epf

The automorphisms
$\ha\sig$
and
$\ha\Sig$
yield the following explicit expressions of the map
$\tau\anl \,.$

\bTh\label{expression of angular parallelisation}
In the non degenerate case, the isomorphism
$\tau\anl$
has the expression
\[
\tau\anl :
T\f S\rot \sub \f S\rot \car \f S\rel \to \f S\rot \car \f V\anl :
(r\rot \,,\, v\rot) \mto
(r\rot \,,\; \Ome) \,,
\]
where
\bEq
\Ome = \ha\sig^{-1}(r\rot) (\sum_i r_i \cro v_i) =
\ha\Sig^{-1}(r\rot) (\sum_i \mu_i \, r_i \cro v_i) \,.
\eEq
\eTh

\bpf
Let
$r\rot \in \f S\rot \,,\;$
$v\rot \eqv (v_1, \dots, v_n) \in T_{r\rot}\f S\rot \sub \f S\rel$
and set
$\Ome \byd \rho\anl (r\rot, v\rot) \in \f V\anl \,.$

The definitions of
$\sig$
and of
$g\rot$
yield, respectively, the following equalities, for each
$\Ome' \in \f V\anl \,,$
\bal
g\rot(r\rot) \, (v\rot \,, \Ome' \cro\mul r\rot)
&\byd
\sig(r\rot) \, (\Ome \,,\, \Ome')
\\
g\rot(r\rot) \, (v\rot \,, \Ome' \cro\mul r\rot)
&\byd
\sum_i g(v_i \,,\, \Ome' \cro r_i) =
g(\Ome' \,,\, \sum_i r_i \cro v_i) =
g(\sum_i r_i \cro v_i \,,\, \Ome') \,.
\end{align*}

Then, by comparison of the above equalities, we obtain
$\sig\Fla\rot (r\rot) (\Ome) =
g\Fla (\sum_i r_i \cro v_i) \,,$
hence
$\ha\sig (r\rot) (\Ome) \byd
(g\Sha \com \sig\Fla\rot) (r\rot) (\Ome) =
\sum_i r_i \cro v_i \,,$
which yields
$\Ome = \ha\sig^{-1}(r\rot) (\sum_i r_i \cro v_i) \,.$

We can prove the 2nd expression of
$\Ome$
in analogous way, by replacing
$g\rot$
with
$G\rot \,.$\QED
\epf

In the classical literature,
$\Ome$
is computed by means of the Poisson's formulas, in terms of a basis.
The above Theorem provides an intrinsic expression of
$\Ome \,,$
which plays an essential role in next sections.

\bNt
The map
$\tau\anl$
is a geometric object, which has nothing to do with masses and
weights, because the rigid constraint does not involve the masses.
Accordingly, the 1st formula in the above Theorem is natural,
while the 2nd one sounds quite strange.
Indeed, the 2nd formula is true for any arbitrary choice of the
weights.
We have added the 2nd formula for the sake of completeness.
Actually, in the 2nd formula, the weights appear both in the
expressions of the sum and of
$\ha\Sig \,;$
eventually, the contribution of the weights disappear.

In order to help understanding this situation, we prove directly that
the 1st formula implies the 2nd one.

Let
$r\rot \in \f S\rot \,,\;$
$v\rot \in T_{r\rot}\f S\rot \sub \f S\rel$
and set
$\Ome \byd \rho\anl (r\rot \,,\, v\rot) \in \f V\anl \,.$

Then, the definition of
$\ha\Sig$
and the 1st expression of
$\rho\anl$
imply
\[
\ha\Sig(r\rot) (\Ome) =
\sum_i \mu_i \, r_i \cro (\Ome \cro r_i) =
\sum_i \mu_i \, r_i \cro
\Big(\big(\ha\sig^{-1}\rot(r\rot)
(\sum_i r_i \cro v_i)\big) \cro r_i\Big)
\,.
\]

On the other hand, in virtue of the definition of
$\rho\anl$
and of its 1st expression, we have
\[
\big(\ha\sig^{-1}\rot (r\rot) (\sum_i r_i \cro v_i)\big) \cro r_i =
v_i \,.
\]

Hence, we obtain
$\ha\Sig(r\rot) (\Ome) =
\sum_i \mu_i \, r_i \cro v_i \,,$
which yields
$\Ome = \ha\Sig^{-1}(r\rot) (\sum_i \mu_i \, r_i \cro v_i) \,.$\END
\eNt

\bCr\label{expression of rotational parallelisation}
In the non degenerate case, the isomorphism
$\tau\rot$
has the equivalent expression
\[
\tau\rot : T\f S\rot \sub \f S\rot \car \f S\rel
\to \f S\rot \car (\f S^* \ten \f S) :
(r\rot \,,\, v\rot) \mto
\Big(r\rot \,,\; g\Sha \big(i(\Ome)\eta\big)\Big) \,,
\]
where
\[
\Ome = \ha\sig^{-1}(r\rot) (\sum_i r_i \cro v_i) =
\ha\Sig^{-1}(r\rot) (\sum_i \mu_i \, r_i \cro v_i) \,.
\]
\eCr

\bpf
It follows from
Proposition \ref{expression of angular parallelisation}
and from the composition of algebraic isomorphisms
\bcd
\f V\anl = \B L^{-1} \ten \f S
@>{i( \cdot)\eta}>>
\B L^2 \ten \Lam^2 \f S^*
@>{g\Sha}>>
\f S^* \ten \f S \,.\QED
\ecd
\epf

\bPr
The eigenvalues of
$\ha\Sig$
turn out to be constant with respect to
$\f S\rot \,,$
in virtue of the invariance of
$\Sig$
with respect to
$SO(\f S, g) \,.$

In the non degenerate case, we have three eigenvalues.
Then, three cases may occur:
\bal
\lam \byd \lam_1 = \lam_2 = \lam_3 \,,
&\qquad
\text{\em spherical case} \,,
\\
\lam \byd \lam_1 = \lam_2 \neq \lam_3 \,,
&\qquad
\text{\em symmetric case} \,,
\\
\lam_1 \neq \lam_2 \neq \lam_3 \neq \lam_1 \,,
&\qquad
\text{\em asymmetric case}.
\end{align*}

In the degenerate case, we have two coinciding eigenvalues
\[
\lam \byd \lam_1 = \lam_2 = \sum_i \mu_i \, g(r_i, r_i) \,.
\]
\ePr

Analogous results hold for
$\ha\sig \,.$

\myskip

We have studied the diagonalisation of
$\Sig$
with respect to
$g \,.$
In an analogous way, we can diagonalise
$G\rot$
with respect to
$G\anl \,.$
Indeed, in this way we obtain the same eigenvalues and the same
classification, because the two diagonalisations are related by the
isomorphism
$\tau\anl \,.$

\myskip

The scaled metric
$m_0 \, \Sig \,,$
or the scaled automorphism
$m_0 \, \ha\Sig \,,$
are called the {\em inertia tensor\/} and the scaled eigenvalues
$I_i = m_0 \, \lam_i : \f S\rot \to (\B L^2 \ten \B M) \ten \Rn$
of the inertia tensor are called {\em principal inertia momenta\/}.
\subsubsection{Continuous interpretation}
\label{Continuous interpretation}
\bsm
We can interpret the above results concerning the parallelisation
of
$\f S\rot$
also in terms of continuous transformations.
Here, in order to keep the thread of our reasoning, we adopt a purely
geometric approach which does not involve time, but this section
can be easily rephrased in a true kinematical way, by replacing
$\Rn$
with
$\f T \,,$
or
$\B T \ten \Rn \,,$
as appropriate.
\esm

We define a \emp{continuous transformation} as a map
\[
C : \Rn \car (\Rn \car \f P) \to \f P \,,
\]
such that, for each
$\tau, \tau', t \in \Rn \,,\; p \in \f P \,,$
\[
C(0,t,p) = p
\ssep{and}
C(\tau', t+\tau, C(\tau, t, p)) = C(\tau+\tau', t, p) \,.
\]

A continuous transformation is said to be \emp{rigid} if, for each
$\tau, t \in \Rn \,,\; p, q \in \f P \,,$
\[
\|C(\tau, t, q) - C(\tau, t, p)\| = \|q - p\| \,.
\]

We can prove that a continuous transformation
$C$
is rigid if and only if its expression is of the type
\[
C(\tau, t, p) = c(t) + \Phi_{(\tau,t)} \, (p - o) \,,
\qquad
\Al \, \tau, t \in \Rn \,,\; p \in \f P \,,
\]
where
$o \in \f P \,,$
$c : \Rn \to \f P$
and
$\Phi : \Rn \car \Rn \to SO(\f S, g) \,.$

Let us suppose that
$C$
be rigid.
The partial derivative of
$\Phi$
with respect to time, at
$\tau = 0 \,,$
turns out to be an antisymmetric endomorphism
\[
\del\Phi : \Rn \to so(\f S, g) \sub \f S^* \ten \f S \,.
\]

Hence, the \emp{velocity} of the continuous transformation
$\E v : \f T \car \f P \to \f S$
is given by
\[
\E v(t, p) = Dc(t) + \del\Phi(t) \, (p - o) \,,
\qquad
\Al \, t \in \f T \,,\; p \in \f P \,.
\]

On the other hand, we obtain the map
\[
\Ome \byd (* \com g\Fla) (\del\Phi) :
\Rn \to \B L^* \ten \f S \,,
\]
via the metric isomorphism
$g\Fla : \f S \to \f S^*$
and the Hodge isomorphism
$* : \Lam^2\f S^* \to \B L^* \ten \f S \,.$

Therefore, we can express the velocity of the continuous
transformation by the classical formula
\[
\E v(t, p) = Dc(t) + \Ome (t) \cro (p - o) \,,
\qquad
\Al \, t \in \Rn \,,\; p \in \f P \,.
\]

\bNt
Let
$\f P\rig \sub \f P_1 \car \dots \f P_n$
be a non degenerate rigid configuration space and
$s\rig : \Rn \to \f P\rig$
be a map.

Then, there is a unique continuous rigid transformation such that the
particles of the continuous transformation, which coincide with the
particles of the discrete rigid system at a certain time, move as the
particles of the discrete rigid system.

In other words, there is a unique rigid continuous transformation
\[
C : \Rn \car (\Rn \car \f P) \to \f P \,,
\]
such that, for each
$p = (p_1, \dots, p_n) \in \f P\rig \,,$
\[
C(\tau, t, p_i) = s_i(t + \tau) \,,
\qquad
\Al \, \tau, t \in \Rn \,.
\]

Then, for each
$p = (p_1, \dots, p_n) \in \f P\rig$
and
$t \in \Rn \,,$
we have
\[
\E v (t, p_i) =
ds_i (t) =
ds\cen (t) + \Ome (t) \cro r_i \,.
\]

Indeed, the rotational components of the velocity of the continuous
and discrete rigid maps coincide.\END
\eNt
\subsection{Splitting of the tangent and cotangent multi--space}
\label{Splitting of the tangent and cotangent multi--space}
\bsm
Next, we study the relation between the tangent and cotangent spaces
of the rigid configuration space and the tangent and
cotangent spaces of the environmental space.
Namely, we exhibit a natural orthogonal splitting of the
environmental tangent and cotangent spaces into three components: the
component of the center of mass, the angular component and the
component orthogonal to the rigid configuration space.
This splitting will have a fundamental role in mechanics of rigid
systems.
\esm
\subsubsection{Splitting of the tangent multi--space}
\label{Splitting of the tangent multi--space}
Let us consider the space
$T\f P\mul|_{\f P\rig} = \f P\rig \car \f S\rel \,.$

\bTh\label{splitting of the tangent rigid space}
We have the orthogonal splitting, with respect to
$G\mul$
\[
T\f P\mul|_{\f P\rig} =
T\f P\rig \udrs{\f P\rig} T\Per \f P\rig =
(T\f P\cen \car T\f S\rot)
\udrs{\f P\rig} T\Per \f P\rig \,,
\]
where
$T\Per\f P\rig$
is the orthogonal complement of
$T\f P\rig$
in
$T\f P\mul|_{\f P\rig} \,.$

The subspace
$T\Per\f P\rig$
is characterised by the following equality
\[
T\Per\f P\rig =
\{ (p\rig \,,\, v\mul) \in \f P\rig \car \f S\mul
\sst
\sum_i \mu_i \, v_i = 0 \,,\; \sum_i \mu_i \, r_i \cro v_i = 0 \}
\sub T\f P\mul|_{\f P\rig} \,.
\]

Moreover, the expressions of the projections associated with the
splitting are
\bAt{2}
T\pi\cen
&: T\f P\mul|_{\f P\rig} \to T\f P\cen
&&:
(p\rig \,,\, v\mul) \mto (p\cen \,,\, v\cen) \,,
\\
T\pi\rot
&: T\f P\mul|_{\f P\rig} \to T\f S\rot
&&:
(p\rig \,,\; v\mul) \mto (r\rot \,,\, \Ome \cro\mul r\rot) \,,
\\
\pi\rig\Per
&: T\f P\mul|_{\f P\rig} \to T\Per\f P\rig
&&:
(p\rig \,,\; v\mul) \mto
(p\rig \,,\; v\mul - v\dia - \Ome \cro\mul r\rot) \,,
\end{alignat}
where
\bgt
p\cen = o + \sum_i \mu_i \, (p_i - o) \,,
\qquad
p\dia \byd (p\cen, \dots, p\cen) \,,
\qquad
r\rot = p\rig - p\dia \,,
\\
v\cen = \sum_i \mu_i \, v_i \,,
\qquad
v\dia \byd (v\cen, \dots, v\cen) \,,
\qquad
v\rot = v\mul - v\dia \,,
\\
\Ome \byd
\ha\Sig^{-1}(r\rot)\big(\sum_i \mu_i \, r_i \cro v_i\big) \,.
\end{gather*}
\eTh

\bpf
The expression of the 1st projection is obvious.

Let us prove the expression of the 2nd projection.
For each
$r\rot \in \f S\rot \,,\; v\rel \in \f S\rel$
and
$\Ome' \in \f V\anl \,,$
in virtue of the definitions of
$\Sig$
and of
$G\rel \,,$
and by a cyclic permutation, we obtain the equalities
\bal
G\rel \, (v\rel \,,\; \Ome' \cro\mul r\rot)
&=
G\rel \,
\big(T\pi\rot(r\rot, v\rel) \,,\; \Ome' \cro\mul r\rot\big) =
G\rel \,
\big(\Ome \cro\mul r\rot \,,\; \Ome' \cro\mul r\rot\big)
\\
&=
\Sig(r\rot) \, \big(\Ome, \Ome'\big) \,,
\\
G\rel \, (v\rel \,, \Ome' \cro\mul r\rot)
&=
\sum_i \mu_i \, g(v_i \,,\; \Ome' \cro r_i) =
g(\Ome' \,,\; \sum_i \mu_i \, r_i \cro v_i) =
g(\sum_i \mu_i \, r_i \cro v_i \,,\; \Ome') \,.
\end{align*}

A comparison of the above equalities yields
$\Sig\Fla(r\rot) \, (\Ome) =
g\Fla(\sum_i \mu_i \, r_i \cro v_i) \,,$
hence
\[
\Ome =
(\Sig\Sha(r\rot) \com g\Fla) (\sum_i \mu_i \, r_i \cro v_i) =
\ha\Sig^{-1}(r\rot) (\sum_i \mu_i \, r_i \cro v_i) \,.
\]

Then, the characterization of
$T\Per\f P\rig$
is easily obtained by considering the multivectors whose previous
projections vanish and by recalling that
$\tau\anl$
and
$\ha\Sig$
are isomorphisms.

Eventually, the 3rd projection is obtained by subtracting
the previous projections.\QED
\epf

We observe that the expression of
$T\pi\rot$
is similar to the 2nd formula of
Proposition \ref{expression of angular parallelisation}.
However, we stress that the multivector
$v\mul$
in the above Theorem needs not to be tangent to
$\f S\rot$
and its projection on
$\f S\rot$
involves the weights.
In the particular case when the multivector
$v\mul$
is tangent to
$\f S\rot \,,$
the expression of
$T\pi\rot$
reduces to the 2nd formula of
Proposition \ref{expression of angular parallelisation}.
\subsubsection{Splitting of the cotangent space}
\label{Splitting of the cotangent space}
Let us consider the space
$T^*\f P\mul|_{\f P\rig} \byd \f P\rig \car \f S^*\rel \,.$

\bTh\label{splitting of the cotangent rigid space}
We have the orthogonal splitting, with respect to
$\ba G\mul$
\[
T^*\f P\mul|_{\f P\rig} =
T^*\f S\rig \udrs{\f P\rig} T^*\per \f P\rig =
(T^*\f P\cen \car T^*\f S\rot) \udrs{\f P\rig} T^*\per \f P\rig \,,
\]
where
$T^*\per \f P\rig$
is the orthogonal complement of
$T^*\f P\rig$
in
$T^*\f P\mul|_{\f P\rig}$
(i.e., the space of annihilators of
$T\f P\rig$).

The subspace
$T^*\per\f P\rig$
is characterised by the following equality
\[
T^*\per\f P\rig =
\big\{ (p\rig \,,\, \alp\mul) \in \f P\rig \car \f S^*\mul
\sst
\sum_i \alp_i = 0 \,,\;
\sum_i g\Fla(v_i) \cro \alp_i = 0 \big\}
\sub T^*\f P\mul|_{\f P\rig} \,.
\]

Moreover, the expressions of the projections associated with the
splitting are
\bAt{2}
T^*\pi\cen
&: T^*\f P\mul|_{\f P\rig} \to T^*\f P\cen
&&:
(p\rig \,,\, \alp\mul) \mto
(p\cen \,,\, \alp\cen)
\\
T^*\pi\rot
&: T^*\f P\mul|_{\f P\rig} \to T^*\f S\rot
&&:
(p\rig \,,\, \alp\mul) \mto (r\rot \,,\, \alp\rot)
\\
\pi\per
&: T^*\f P\mul|_{\f P\rig} \to T^*\per\f P\rig
&&:
(p\rig \,,\, \alp\mul) \mto
(r\rot \,,\, \alp\mul - \alp\dia - \alp\rot) \,,
\end{alignat}
where
\bGt
p\cen = o + \sum_i \mu_i \, (p_i - o) \,,
\qquad
p\dia \byd (p\cen, \dots, p\cen) \,,
\qquad
r\rot = p\rig - p\dia \,,
\nonumber
\\
\alp\cen = \sum_i \alp_i \,,
\qquad
\alp\dia = (\mu_1 \, \alp\cen, \dots, \mu_n \, \alp\cen) \,,
\nonumber
\\
\alp\rot =
\Big((\ha\Sig^{-1})^*(r\rot)
\big(\sum_i g\Fla(r_i) \cro \alp_i\big)\Big)
\cro\mul G\mul\Fla(r\rot) \,.
\end{gather}
\eTh

\bpf
The commutative diagram
\bcd
\B L^2 \ten T\f P\mul|_{\f P\rig} @>{T\pi\rot}>> \B L^2 \ten T\f S\rot
\\
@V{G\Fla\mul}VV @VV{G\Fla\rot}V
\\
T^*\f P\mul|_{\f P\rig} @>{T^*\pi\rot}>> T^*\f S\rot
\ecd
Theorem \ref{splitting of the tangent rigid space}
and the definition of
$\ha\Sig$
give
\bal
T^*\pi\rot(p\rig \,,\, \alp\mul)
&=
(G\rot\Fla \com T\pi\rot \com G\mul\Sha) (p\rig \,,\, \alp\mul)
\\
&=
(G\rot\Fla \com T\pi\rot) \big(p\rig \,,\;
\tfr1{\mu_1} g\Sha(\alp_1),
\dots,
\tfr1{\mu_n} g\Sha(\alp_n)\big)
\\
&=
G\rot\Fla \Big(r\rot \,,\;
\ha\Sig^{-1}(r\rot)
\big(\sum_i r_i \cro g\Sha(\alp_i)\big) \cro\mul r\rot\Big)
\\
&=\Big(r\rot \,,\;
\big((g\Fla \com \Sig\Sha)(r\rot)
(\sum_i g\Fla(r_i) \cro \alp_i)\big)
\cro\mul G\mul\Fla(r\rot)\Big)
\\
&=\Big(r\rot \,,\;
\big((\ha\Sig^{-1})^*(r\rot)
(\sum_i g\Fla(r_i) \cro \alp_i)\big)
\cro\mul G\mul\Fla(r\rot)\Big) \,.\QED
\end{align*}
\epf

The projection
$T^*\pi\rot$
can be expressed in terms of
$\f V\anl \,.$
In this way, we recover the classical formula of the ``total momentum"
of a multi--form.
Here, this formula arises naturally from our geometric interpretation
of
$T^*\f S\rot \,.$

\bCr\label{total momentum of a multiform}
We have the projection
\bMl
\C S\anl \byd (\tau\anl^{-1})^* \com T^*\pi\rot :
T^*\f P\mul|_{\f P\rig} \to \f S\rot \car \f V\anl^* :
\\
: (p\rig, \, \alp\mul) \mto
\big(r\rot,\; \sum_i g\Fla(r_i) \cro \alp_i\big) \,,
\end{multline}
where
$r\rot \byd \pi\rot(p\rig) \,.$
\eCr

\bpf
By recalling the expressions of
$T^*\pi\rot \,,\;$
$(\ha\Sig^{-1})^* \,,\;$
$(\tau\anl^{-1})^*$
and
$\ha\Sig \,,$
we obtain
\bal
\C S\anl (p\rig, \, \alp\mul)
&=
\big((\tau\anl^{-1})^* \com T^*\pi\rot\big) (p\rig, \, \alp\mul)
\\
&=
(\tau\anl^{-1})^*\Big(r\rot, \;
\big((\ha\Sig^{-1})^* (r\rot)
(\sum_j g\Fla(r_j) \cro \alp_j)\big)
\cro\mul G\mul\Fla(r\rot)\Big)
\\
&=
\Big(r\rot,\; \sum_i g\Fla(r_i) \cro
\big(\big((\ha\Sig^{-1})^* (r\rot)
(\sum_j g\Fla(r_j) \cro \alp_j)\big)
\cro \mu_i \, g\Fla(r_i)\big)\Big)
\\
&=
\Big(r\rot,\; \sum_i \Big(g\Fla(r_i) \cro\mul G\mul\Fla
\big(\ha\Sig^{-1}(r\rot) (\sum_j r_j \cro g\Sha(\alp_j)))
\cro r\rot\big)_i\Big)\Big)
\\
&=
\Big(r\rot, \; g\Fla
\big(\sum_i \mu_i \, r_i \cro \big(\ha\Sig^{-1}(r\rot)(\sum_j \,
r_j \cro g\Sha(\alp_i)) \cro r_i
\big)\big)\Big)
\\
&=
\big(r\rot, \; \sum_i g\Fla(r_i) \cro \alp_i\big) \,.\QED
\end{align*}
\epf
\subsection{Kinetic energy and momentum of the rigid system}
\label{Kinetic energy and momentum of the rigid system}
According to our scheme, we define the \emp{rigid kinetic energy},
the \emp{rigid kinetic momentum}, the \emp{rotational kinetic energy}
and the \emp{rotational kinetic momentum} as
\bat{3}
\C K\rig
&\byd
i\rig^* \C K\mul
&&:
\B T^{-1} \ten T\f P\rig \to
(\B T^{-2} \ten \B L^2 \ten \B M) \ten \Rn
&&:
v\rig \mto  \tfr12 m_0 \, G\rig(v\rig,v\rig) \,,
\\
\C P\rig
&\byd
i\rig^* \C P\mul
&&:
\B T^{-1} \ten T\f P\rig \to
(\B T^{-1} \ten \B L^2 \ten \B M) \ten T^*\f P\rig
&&:
v\rig \mto m_0 \, G\rig\Fla(v\rig) \,,
\\
\C K\rot
&\byd
i\rot^* \C K\mul
&&:
\B T^{-1} \ten T\f S\rot \to
(\B T^{-2} \ten \B L^2 \ten \B M) \ten \Rn
&&:
v\rot \mto  \tfr12 m_0 \, G\rot(v\rot,v\rot) \,,
\\
\C P\rot
&\byd
i\rot^* \C P\mul
&&:
\B T^{-1} \ten T\f S\rot \to
(\B T^{-1} \ten \B L^2 \ten \B M) \ten T^*\f S\rot
&&:
v\rot \mto m_0 \, G\rot\Fla(v\rot) \,.
\end{alignat*}

Then, by taking into account the angular parallelisation
$\tau\anl : T\f S\rot \to \f S\rot \car \f V\anl \,,$
we obtain the \emp{angular kinetic energy} and the
\emp{angular kinetic momentum}
\bat{2}
\C K\anl
&\byd
\tau\anl^* \, \C K\rot : \B T^{-1} \ten T\f S\rot \to
(\B T^{-2} \ten \B L^2 \ten \B M) \ten \Rn
\\
&:
(r\rot, v\rot) \mto \tfr12 m_0 \, \Sig(r\rot)(\Ome,\Ome)
= \tfr12 m_0 \, \sum_i \mu_i \, \big(
g(r_i, r_i) \, g(\Ome, \Ome) -
g(r_i, \Ome) \, g(r_i, \Ome)
\big) \,,
\\
\C P\anl
&\byd
\tau\anl^* \, \C P\rot :
\B T^{-1} \ten T\f S\rot \to
(\B T^{-1} \ten \B L^2 \ten \B M) \ten \f V\anl^*
\\
&:
(r\rot, v\rot) \mto m_0 \, \Sig\Fla(r\rot)(\Ome) =
m_0 \, \sum_i \mu_i \, \big(
g(r_i, r_i) \, g\Fla(\Ome) -
g(r_i, \Ome) \, g\Fla(r_i)
\big ) \,,
\end{alignat*}
where
$\Ome \byd \rho\anl(v\rot) \,.$

According to the splittings
$T\f P\rig = T\f P\cen \car T\f S\rot$
and
$T\f P\rig = T\f P\cen \car (\f S\rot \car \f V\anl) \,,$
we have, respectively, the splittings
\bat{4}
\C K\rig
&= \C K\cen + \C K\rot \,,
\qquad
\C P\rig
&&=
(\C P\cen, \, \C P\rot) \,,
\\
\C K\rig
&= \C K\cen + \C K\anl \,,
\qquad
\C P\rig
&&=
(\C P\cen, \, \C P\anl) \,.
\end{alignat*}
\subsection{Connections induced on the rigid system}
\label{Connections induced on the rigid system}
First of all, in the non degenerate case, the generalised affine
structure of
$\f P\rig$
induces a flat connection
$\nab_{\aff}$
(see Section \ref{Generalised affine spaces}).

Moreover, according to the Gauss' Theorem (see, for instance,
\cite{GalHulLaf90}), the geometric metric
$g\rig$
and the weighted metric
$G\rig$
induce two distinct connections
$\nab\rig$
and
$\nab\Rig \,,$
respectively,
on
$\f P\rig \,.$
Actually, we shall be mainly concerned with
$\nab\Rig \,,$
which is the most important of the two, because of its role in
dynamics.

\bPr
The connection
$\nab\Rig$
splits into the cartesian product of the connections
$\nab\cen$
and
$\nab\rot$
of
$\f P\cen$
and
$\f S\rot \,.$\END
\ePr
\subsection{Kinematics of rigid systems}
\label{Kinematics of rigid systems}
\bsm
In this section, we apply the splitting of
$T\f P\mul|_{\f P\rig}$
to the velocity and the acceleration of a rigid system.
Namely, the velocity splits into the two components of the center of
mass and the velocity relative to the center of mass.
On the other hand, the acceleration splits into the three components
of the center of mass, relative to the center of mass and the term
given by the 2nd fundamental form of the rigid configuration space.
\esm

Let us consider a rigid motion
$s\rig : \f T \to \f P\rig \,.$

\bPr
According to
Corollary \ref{angular parallelisation}
and
Theorem \ref{expression of angular parallelisation},
we obtain the splittings
\bat{2}
s\rig
&= (s\cen,s\rot)
&&:
\f T \to \f P\cen \car \f S\rot
\\
ds\rig
&= \big(ds\cen, \, (s\rot,\Ome)\big)
&&:
\f T \to (\B T^{-1} \ten T\f P\cen) \car
\big(\f S\rot \car (\B T^{-1} \ten \f V\anl)\big) \,,
\end{alignat*}
where
\[
s\cen \byd \pi\cen \com s\rig \,,
\qquad
s\rot \byd \pi\rot \com s\rig \,,
\qquad
\Ome \byd \ha\Sig^{-1}(\sum_i \mu_i \, (s\rot)_i \cro d(s\rot)_i)
\,.\END
\]
\ePr

The map
$\Ome \byd \rho\anl \com T\tau\anl \com ds\rig :
\f T \to \B T^* \ten \f V\anl$
is called the {\em angular velocity\/} of the rigid motion.

\bTh\label{splitting of the acceleration}
The acceleration splits into the three components as
\[
\nab\mul ds\rig = \nab\cen ds\cen + \nab\rot ds\rot + N \com ds\rot
\,,
\]
where
$N : T\f P\rig \to T\Per\f P\rig$
is the 2nd fundamental form of the connection
$\nab\mul \,,$
with respect to the metric
$G\mul \,.$
We have the expressions
\bGt
\rho\anl(\nab\rot ds\rot) =
d\Ome +
\ha\Sig^{-1}(s\rot)\big(\Ome \cro \ha\Sig(s\rot)(\Ome)\big)
\,,
\\
N (s\rig)(\Ome) =
\Ome \cro\mul (\Ome \cro\mul s\rot) -
\ha\Sig^{-1}(s\rot)
\big(\Ome \cro \ha\Sig(s\rot)(\Ome)\big) \cro\mul s\rot \,.
\end{gather}
\eTh

\bpf
The proof is analogous to that for the case of one constrained
particle.

In virtue of
Theorem \ref{expression of angular parallelisation},
Corollary \ref{angular parallelisation}
and the Leibnitz identity for the cross product,
we obtain
\begin{gather*}
\rho\anl(\nab\rot ds\rot) =
\\
\begin{align*}
&=
(\rho\anl \com T\pi\rot) (\nab\mul ds\rot)
\\
&=
\ha\Sig^{-1}(s\rot)
\big(\sum_i \mu_i (s\rot)_i \cro \nab d(s\rot)_i\big)
\\
&=
\ha\Sig^{-1}(s\rot)
\big(\sum_i \mu_i (s\rot)_i
\cro \nab (\Ome \cro (s\rot)_i))
\\[2mm]
& = \ha\Sig^{-1}(s\rot)
\big(\sum_i \mu_i (s\rot)_i \cro (d\Ome \cro (s\rot)_i)\big) +
\ha\Sig^{-1}(s\rot)
\big(\sum_i \mu_i (s\rot)_i \cro (\Ome \cro (\Ome \cro
(s\rot)_i))\big)
\\
&=
\ha\Sig^{-1}(s\rot)
\big(\sum_i \mu_i (s\rot)_i \cro (d\Ome \cro (s\rot)_i)\big) +
\ha\Sig^{-1}(s\rot)
\big(\sum_i \mu_i \Ome \cro ((s\rot)_i \cro (\Ome \cro(s\rot)_i)\big)
\\
&=
d\Ome +
\ha\Sig^{-1}(s\rot)\big(\Ome \cro \ha\Sig(s\rot)(\Ome)\big)
\,.
\end{align*}
\end{gather*}

Moreover, we have
\begin{gather*}
N(ds\rot) =
\\
\begin{align*}
&=
\nab\mul ds\rot - \nab\rot ds\rot
\\
&=
d\Ome \cro\mul s\rot + \Ome \cro\mul (\Ome \cro\mul s\rot) -
d\Ome \cro\mul s\rot -
\ha\Sig^{-1}(s\rot)\big(\Ome \cro \ha\Sig(s\rot)(\Ome)\big)
\cro\mul s\rot
\\
&=
\Ome \cro (\Ome \cro\mul s\rot) -
\ha\Sig^{-1}(s\rot)
\big(\Ome \cro \ha\Sig(s\rot) (\Ome)\big) \cro\mul s\rot \,.\QED
\end{align*}
\end{gather*}
\epf

\bNt
The map
$d\Ome$
is the covariant derivative of
$\Ome$
with respect to the natural connection
$\nab_{\aff}$
of
$\f S\rot$
induced by
$\tau\anl \,.$
Hence, the map
$\ha\Sig^{-1}(s\rot) \big(\Ome \cro \ha\Sig(s\rot)(\Ome)\big)$
expresses the Christoffel symbol of the connection
$\nab\rot$
with respect to the parallelisation
$\tau\anl \,.$\END
\eNt
\subsection{Dynamics of rigid systems}
\label{Dynamics of rigid systems}
\bsm
In this section we study the equation of motion for a rigid system.

According to the results of the previous section, we show that the
equation of motion in the environmental space splits into three
components: the equation of motion for the center of mass (related to
the linear momentum), the equation of motion for the relative motion
(related to the angular momentum) and equation for the reaction force
(related to the 2nd fundamental form of the rigid configuration
space).
\esm
\subsubsection{Splitting of multi-forces}
\label{Splitting of multi-forces}
According to the scheme discussed for a constrained system, we assume
a multi-force
\[
\ti F\mul :
J_1\f P\mul \to (\B T^{-2} \ten \B L^2 \ten \B M) \ten T^*\f P\mul
\]
and consider its restriction to the phase space of the rigid system
\[
F \byd \ti F\mul|_{J_1\f P\rig} :
J_1\f P\rig \to (\B T^{-2} \ten \B L^2 \ten \B M) \ten T^*\f P\mul \,.
\]

\bPr
According to the splitting of
$T^*\f P\mul|\f P\rig \,,$
we can write
\[
F = F\rig + F\rig\,\per \,,
\]
where
\[
F\rig = i\rig^* F :
J_1\f P\rig \to (\B T^{-2} \ten \B L^2 \ten \B M) \ten T^*\f P \,.
\]

Moreover, according to the splitting of
$T^*\f P\rig$
(see Theorem \ref{splitting of the cotangent rigid space}),
we can write
\[
F\rig = (F\cen, F\rot) = (F\cen, F\anl) \,,
\]
where
$F\cen$
and
$F\anl$
turn out to be, respectively, the \emp{total force} and the \emp{total
momentum of the force}.
We have the following expressions
\bgt
F\cen =
T^*\pi\cen \com F\rig =
\sum_i F_i \,,
\qquad
F\anl =
\C S\anl \com F\rig = \sum_i \bau r_i \cro F_i \,,
\\
F\rot =
((\Sig^{-1})^* (F\anl)) \cro\mul \bau r\rig \,,
\qquad
F\dia = (\mu_1 \, F\cen, \, \dots, \, \mu_n \, F\cen) \,,
\\
F\per =
F\mul - F\dia - F\rot \,,
\end{gather*}
where we have set
$\bau r\rig :\f S\rot \sub \f S\rel \to \B L^2 \ten \f S^*\rel :
r\rot \mto G\mul\Fla(r\rot) \,.$\END
\ePr

\bNt
If the multi-force
$F\mul$
fulfills the 3rd Newton's principle then its
component tangent to the constraint
$F\rig$
vanishes.
\eNt

\bpf
If
$F\mul$
fulfills the 3rd Newton's principle, then we obtain
$F\cen = 0$
and
$F\anl = 0 \,,$
hence
$F\rig = 0 \,.$\QED
\epf

Moreover, we assume a reaction force
\[
R :
J_1\f P\rig \to \B T^{-2} \ten \B L^2 \ten \B M \ten T^*\f P\mul \,,
\]
which splits analogously to the force.

The above Note holds also for the reaction.
So, we assume
$R\rig = 0 \,,$
i.e.
$R = R\per \,.$
\subsubsection{Splitting of the equation of motion}
\label{Splitting of the equation of motion}
Eventually, we are ready to split the equation of motion into three
components. We follow the scheme developed for one constrained
particle with the additional results arising from the present
framework.

\bCr
The rigid motion
$s\rig : \f T \to \f P\rig$
and the reaction force
$R$
fulfill the Newton's law of motion
\[
m_0 \, G\mul\Fla (\nab\mul ds\rig) =
(F\rig + R) \com j_1s\rig
\]
if and only if
\bGt
m_0 \, G\cen\Fla (\nab\cen ds\cen) =
F\cen \com j_1s\rig \,,
\qquad
m_0 \, \ha\Sig \big(\tau\anl(\nab\rot ds\rot)\big) =
F\anl \com j_1s\rig
\\
m_0 \, G\mul\Fla (N \com ds\rot) =
(F\per + R\per) \com j_1s\rig \,,
\end{gather}
where
\[
\rho\anl(\nab\rot ds\rot) = d\Ome +
\ha\Sig^{-1}(s\rot)
\big(\Ome \cro \ha\Sig(s\rot)(\Ome)\big) \,.
\]
\eCr

\bpf
The proof follows from
Theorem \ref{splitting of the acceleration}.\QED
\epf

The angular component of the above equation of motion is referred to
as \emp{Euler's equation\/}.

\bNt
Contrary to the general constrained case, here the equation on
the constrained space can be split into the center of mass and
rotational components, due to the fact that the rigid constraint is
without interference between particles and the center of mass.

On the other hand, as in the case of a system of free particles, we
cannot solve the first two equations independently, unless the total
force factors through the projection
$\pi\cen$
on
$\f P\cen \,.$\END
\eNt

\bCr\label{expression of the rigid reaction}
The reaction
$R\per$
is given by
\bMl
R\per =
\bau\Ome \cro \big(\bau\Ome \cro \bau r\rig\big) -
\Sig^{-1}{}^* (s\rot)
\Big(
\big(\bau\Ome \cro \Sig\Fla(s\rot)(\Ome)\big) \cro\mul \bau r\rig
\Big)
\\
- F\mul + (\mu_1 \, F\cen, \, \dots, \, \mu_n \, F\cen) +
\big((\Sig^{-1})^* (F\anl)\big) \cro\mul \bau r\rig \,.
\end{multline}
where we set
$\bau\Ome \byd g\Fla(\Ome) \,.$
\eCr

\bpf
The reaction
$R\per$
is determined by on solutions of the equations of motion by the
following equalities
\bal
R\per
&=
G\mul\Fla (N \com ds\rot) - F\per \com j_1s\rig
\\
&=
G\mul\Fla (N \com ds\rot) - (F\mul - F\dia - F\rot) \com j_1s\rig
\\
&=
G\mul\Fla\big(s\rig,
\Ome \cro\mul (\Ome \cro\mul s\rot) -
\ha\Sig^{-1}(s\rot)\big(\Ome \cro \ha\Sig(s\rot)(\Ome)\big)
\cro\mul s\rot
\big)
\\
&\hphantom{=\ }
- F\mul \com j_1s\rig
+ (\mu_1 \, F\cen, \, \dots, \, \mu_n \, F\cen) \com j_1s\rig
+
\Big(g\Fla \big(\Sig\Sha(F\anl)\big)\Big) \com j_1s\rig
\cro\mul \bau s\rig
\\
&=
\big(s\rig, \; \Ome \cro (\bau\Ome \cro\mul \bau s\rig) -
g\Fla \com \Sig\Sha (s\rot) (\bau\Ome \cro
\Sig\Fla(s\rot)(\Ome)) \cro\mul \bau s\rig \big)
\\
&\hphantom{=\ }
- F\mul \com j_1s\rig
+ (\mu_1 \, F\cen, \, \dots, \, \mu_n \, F\cen) \com j_1s\rig
+
\Big(g\Fla\big(\Sig\Sha(F\anl)\big)\Big) \com j_1s\rig
\cro\mul \bau s\rig \,.\QED
\end{align*}
\epf

\myskip

Now, we express the Newton's law in Lagrangian form, in our
special case of rigid systems.
To this aim, we introduce an appropriate chart on
$\f P\rig \,.$
We refer to a chart
$(x^i)$
on
$\f P\cen$
and to a chart (for instance, the \emp{Euler's angles})
$(\alp^j)$
on
$\f S\rot \,.$
Then, the induced chart on
$T\f{Q}$
is
$(x^i, \alp^j, \dt x^i, \dt \alp^j) \,.$

Suppose that
$F\rig :
\f P\rig \to (\B T^{-2} \ten \B L^2 \ten \B M) \ten T^*\f P\rig$
be a conservative positional force, with potential
$\C U\rig :
\f P\rig \to (\B T^{-2} \ten \B L^2 \ten \B M) \ten \Rn \,,$
and that
$R\rig = 0 \,.$
Then, the induced Lagrangian function turns out to be the map
\[
\C L\rig \byd \C K\rig - \C U\rig :
T\f P\rig \to (\B T^{-2} \ten \B L^2 \ten \B M) \ten \Rn \,.
\]

\bCr
Let
$s\rig : \f T \to \f P\rig$
be a motion.
Then,
$s\rig$
and the reaction force
$R$
fulfill the Newton's law of motion if and only if the following
equations hold
\bgt
D\left(\frac{\der \C L\rig}{\der \dot x^i} \com ds\rig\right) -
\frac{\der\C L\rig}{\der x^i} \com ds\rig = 0 \,,
\qquad
D\left(\frac{\der \C L\rig}{\der \dot{\alp}^i} \com ds\rig\right ) -
\frac{\der \C L\rig}{\der \alp^i} \com ds\rig = 0 \,,
\\
G\mul\Fla (N \com ds\rig) = (F\per + R\per) \com j_1s\rig \,.\END
\end{gather*}
\eCr

\end{document}